\documentclass[journal]{IEEEtran}

\usepackage{amsmath,amssymb,amsfonts}
\usepackage{mathtools}
\usepackage{amsthm}
\usepackage{xcolor}
\usepackage{cite}
\usepackage{graphicx}
\usepackage{booktabs}
\usepackage{url}
\usepackage{algorithm}
\usepackage{float}
\newtheorem{assumption}{Assumption}
\newtheorem{lemma}{Lemma}
\newtheorem{theorem}{Theorem}
\newtheorem{corollary}{Corollary}
\newtheorem{proposition}{Proposition}
\newtheorem{remark}{Remark}

\newcommand{\R}{\mathbb R}
\newcommand{\E}{\mathbb E}
\newcommand{\Prob}{\mathbb P}
\newcommand{\tr}{\operatorname{tr}}

\newcommand{\rank}{\operatorname{rank}}
\newcommand{\diag}{\operatorname{diag}}
\newcommand{\col}{\operatorname{col}}
\newcommand{\lammax}{\lambda_{\max}}
\newcommand{\lammin}{\lambda_{\min}}

\newcommand{\new}[1]{#1}          

\newcommand{\sym}{\operatorname{sym}}

\makeatletter
\let\origthebibliography\thebibliography
\renewcommand{\thebibliography}[1]{\origthebibliography{#1}\scriptsize}
\makeatother

\begin{document}
\setlength{\dbltextfloatsep}{6pt plus 1pt minus 1pt}
\setlength{\textfloatsep}{6pt plus 1pt minus 1pt}

\title{  Data-Driven LQR with Uncertainty Refinement and
Domain Randomization---Part~I: Certified Uncertainty
Refinement}

\author{Hamidreza Modares %
\thanks{H. Modares is with Michigan State University,
East Lansing, MI 48824 USA.}}

\maketitle


\begin{abstract}
Data-driven control from trajectories corrupted by both process and measurement noise is challenging when the system matrices and noise covariances are all unknown. Measurement noise correlates the regressor with the noise, biasing standard estimators, and quantifying model uncertainty requires the noise covariance, which can itself only be inferred from the uncertain model. This paper, Part~I of a two-part study, develops a finite-sample framework that quantifies, and then iteratively refines, certified uncertainty sets for the model, the realized noise, and the noise covariance. We derive a closed-loop parametrization that separates the realized noise, i.e., the fixed but unknown noise values in the collected data, from future disturbances, which are random and characterized by their covariance. Delayed instrumental variables (IVs) remove the bias from the regressor–noise correlation and yield valid initial uncertainty sets. {Every candidate model reproduces the data with exactly one realized-noise sequence, so model and noise uncertainties are refined together, and a model is removed only when its noise sequence violates the prior or the statistical certificates.} This correspondence enables a coupled recursion: the noise sequences of the surviving models yield a tighter full-matrix covariance certificate, and each tighter certificate removes further models and their noise sequences. We prove finite-sample containment of the true model, realized noise, and covariance, together with recursive nesting of the certificates. {Several delayed instruments make the certificates overidentifying, which} {tightens the model set and} {yields a certified test of data consistency.} The model and normalized-noise descriptions converge at the finite-sample IV rate, and, under a strict initialization-margin condition, so does the covariance certificate. Part~II uses these certificates for domain-randomized linear quadratic regulator synthesis.
\end{abstract}


\section{Introduction}
\IEEEPARstart{D}{ata-driven} control seeks to synthesize controllers from measured trajectories when accurate models are unavailable. Such approaches are either indirect, identifying a model before synthesis, or direct, bypassing explicit identification~\cite{9705109,DirIndir1}. A natural benchmark for testing these data-driven approaches is the linear quadratic regulator (LQR): its model-based theory is well established, so the gap to the known-model solution can be measured in performance, robustness, and feasibility. We therefore focus our review on data-driven LQR, although the certificates developed here do not depend on the control objective.

Existing direct approaches to the LQR problem include reinforcement-learning methods~\cite{LQR1,LQR2,LQR3,samplecomp,sforni2025stabilitycertifiedonpolicydatadrivenlqr,Zhao2025DeePO,Ma2025RobustLQR} and data-driven closed-loop parametrizations~\cite{DEPERSIS2021109548,ESMZAD2025112197,8960476,Direct2,DAI20231376,Direct3,8933093,Floriancov1,babazadeh,DUST,9992774,Direct1,ANDERSON2019364}. However, their guarantees typically rely on clean state measurements. On the other hand, formulations that account for measurement noise~\cite{10549788,10124991,Data-new1,Data-new2,Data-new3,ramin,Wang2026LQG,out} assume known noise covariances or prescribed noise bounds. Besides, they often suppress the effect of the unknown noise realizations through regularization, which can bias the controller. A challenge is that measurement noise correlates the regressor with the noise, so least-squares estimates remain biased even with abundant data.

Indirect methods face the same obstacle. Finite-sample and certainty-equivalence analyses of LQR~\cite{Simchowitz2018Learning,Dean2020CoarseID,Mania2019CE} rely on quantified identification error, which is only available when the noise statistics or bounds on them are known. Beyond the LQR context, identification methods such as set-membership identification can combine prior information with data to construct sets of admissible models or disturbances~\cite{UncertaintyLearning2,alanwar2023data,Thorpe2024Prior,Alanwar2022SideInfo}. However, they still require prescribed disturbance bounds.  The remaining challenge is twofold: to quantify model uncertainty directly from data when both the system matrices and the noise covariances are unknown (this paper, Part~I), and to use it for controllers that are less conservative than worst-case robust designs, through domain-randomized synthesis (Part~II~\cite{PartII}). The main contributions of Part~I are:
\begin{itemize}
\item {\emph{Coupled model--noise uncertainty.} Delayed IVs embedded in a closed-loop parametrization remove the regressor--noise bias and separate the realized noise in the data from future disturbances, so model uncertainty is expressed through, and quantified jointly with, the unknown noise covariance.}
\item {\emph{Consistency-based refinement and testing.} Models whose implied noise violates the prior or the statistical moment conditions are discarded. With several instruments these conditions are overidentifying and give a certified test of data consistency at no extra confidence cost.}
\item {\emph{Covariance certification without known noise statistics.} A coupled recursion certifies a full-matrix covariance bound that contracts the model and noise sets, with finite-sample containment, recursive nesting, convergence at the IV rate, and a recursion-free self-consistent model set.}
\item {\emph{Quantified gain and separate noise bounds.} The refinement gain is characterized per direction by a support radius, and the process- and measurement-noise covariances are bounded separately when the refined set certifies that $A^\star$ is nonsingular, as Part~II requires.}
\end{itemize}

\textit{Notation:} $\mathbb S^n_+$ is the cone of symmetric positive-semidefinite matrices, and $X\succeq Y$ ($X\succ Y$) means $X-Y\in\mathbb S^n_+$ (positive definite). $X^\dagger$, $\mathrm{row}(X)$ and $X^\perp$ denote the pseudoinverse, row space and a basis of $\ker(X)$. $X^{1/2}$ is the symmetric square root of $X\succeq0$. $\lambda_{\min}$, $\lambda_{\max}$ and $\sigma_{\min}$ are extreme eigenvalues and the smallest singular value. $\|\cdot\|_2$ is the Euclidean or spectral norm, $\|\cdot\|_F$ the Frobenius norm and $\|X\|_*:=\sum_i\sigma_i(X)$ the nuclear norm. $\langle X,Y\rangle:=\mathrm{tr}(X^\top Y)$. $\mathrm{col}(\cdot)$ is vertical concatenation, $[\tilde w_k]_{k\in\mathcal I}$ horizontal concatenation in increasing $k$, $|\mathcal I|$ cardinality, and $\mathrm{vert}(\cdot)$ the vertex set. In a symmetric matrix, $\star$ is the transposed off-diagonal block, and as a superscript, it marks true quantities. $[a]_+:=\max\{a,0\}$, and $k\equiv q\pmod r$ means $r$ divides $k-q$. $S_1\in\mathbb R^{T\times T}$ is the lag-one shift, $[S_1]_{k,k+1}=1$ {with all other entries zero}.

\section{Problem Formulation and Standing Assumptions}
\label{sec:problem}
\subsection{System Model and Assumptions}
Consider a linear time-invariant system
\begin{align}
    x_{k+1} = A^\star x_k + B^\star u_k + \omega_k,  \quad
    y_k &= x_k + \upsilon_k, \label{eq:syst}
\end{align}
where $x_k\in\mathbb{R}^n$ is the state, $u_k\in\mathbb{R}^m$ is the control input,
$y_k\in\mathbb{R}^n$ is the noisy state measurement, and $\omega_k$
and $\upsilon_k$ are the process and measurement noise.

\begin{assumption}
\label{noises}
The sequences \(\{\omega_k\}\) and \(\{\upsilon_k\}\) are i.i.d.\ Gaussian,
\(
\omega_k\sim\mathcal N(0,W),\,
\upsilon_k\sim\mathcal N(0,V),
\)
mutually independent, with unknown covariances
\(W\succ 0\) and $V \succeq 0$.
\end{assumption}

\begin{assumption}
\label{ass:CMZ-prior}
The matrices
\(
\theta^\star:=[A^\star\ B^\star]\in\mathbb R^{n\times (n+m)}
\)
are unknown, but belong to the known matrix zonotope
\begin{align}
\mathcal M_{\rm \theta,p}
:=
\Big\{
C_{\rm \theta,p}
+
\sum_{i=1}^{{q_{\theta,p}}}
G_{{\rm \theta, p},i}\zeta_i:
\|\zeta\|_\infty\leq1
\Big\}.
\label{eq:CMZ-prior}
\end{align}
\end{assumption}

Substituting $x_k=y_k-\upsilon_k$ into \eqref{eq:syst} and using $y_{k+1}=x_{k+1}+\upsilon_{k+1}$ gives
\begin{equation}
y_{k+1} = A^\star y_k + B^\star u_k + \omega_k + \upsilon_{k+1} - A^\star \upsilon_k. \label{eq:y_dyn}
\end{equation}
With the effective noise
\begin{equation} \label{effective1}
{\tilde w_k} := \omega_k + \upsilon_{k+1} - A^\star \upsilon_k,
\end{equation}
equation \eqref{eq:y_dyn} becomes
\begin{equation}
y_{k+1} =  A^\star y_k + B^\star u_k+\tilde w_k,
\label{eq:effective_disturbance}
\end{equation}
and, under Assumption~\ref{noises},
\begin{align} \label{sigmatrue}
 \mathbb{E}[\tilde w_k \tilde w_k^\top] =  \Sigma_{\rm true}
=
W+V+A^\star V(A^\star)^\top.
\end{align}
Since $\tilde w_k$ and $y_k$ share $\upsilon_k$,
$\mathbb E[\tilde w_k y_k^\top]=-A^\star V\neq0$ in general, so a regression on the current measured state inherits a noise correlation.

We organize the collected trajectory of length $T$ as
\begin{subequations}
\label{eq:collected_data}
\begin{align}\nonumber
U_0 &:=
\begin{bmatrix}
u_0 & \cdots & u_{T-1}
\end{bmatrix},\quad
Y_0 :=
\begin{bmatrix}
y_0 & \cdots & y_{T-1}
\end{bmatrix},\\
Y_1 &:=
\begin{bmatrix}
y_1 & \cdots & y_T
\end{bmatrix},
\end{align}
\end{subequations}
with regressor
\begin{align}
\phi_k:=[\,y_k^\top\;\;u_k^\top\,]^\top,
\qquad
\Phi:=
\begin{bmatrix}
\phi_0 & \cdots & \phi_{T-1}
\end{bmatrix}.
\label{phi1}
\end{align}
To obtain a statistically centered relation, we replace the current measured state by its delayed value in the instrumental variable (e.g., \cite{zhang2025error})
\begin{align}
\psi_k:=[\,y_{k-1}^\top\;\;u_k^\top\,]^\top, \qquad \Psi:=\begin{bmatrix}
\psi_0 & \cdots & \psi_{T-1}
\end{bmatrix}
\end{align} 
and, more generally, $N_\psi\ge1$ instruments $\psi_k^{(i)}\in\R^{n+m}$, $i=1,\ldots,N_\psi$, {fixed before the data are processed, with $\psi_k^{(1)}=\psi_k$. Each stacks past measurements, past inputs or exogenous excitation, so that it is uncorrelated with $\tilde w_k$. Instruments built from different signals see the realized noise through different projections, so their certificates intersect in a smaller set and fail differently when the model class is wrong. For $m=1$, a natural second instrument is the past-input vector $\psi_k^{(2)}=[u_{k-1}^\top\,\cdots\,u_{k-n}^\top\;u_k^\top]^\top$. We call $\psi^{(1)}$ the primary instrument; its exogeneity and rank must be checked for the actual data-collection policy. If the instruments use lags up to $L\ge1$, recording starts $L$ samples before $k=0$, and these presample values enter only the first $L$ columns of each $\Psi^{(i)}$.} With
\(
\Psi^{(i)}:=
[
\psi_0^{(i)} \quad ... \quad \psi_{T-1}^{(i)}
]\)
and $\Psi:=\Psi^{(1)}$, we form the IV-normalized data matrices
\begin{align}
&\bar U_0
:=
\tfrac{1}{T}U_0\Psi^\top,\,
\bar Y_0
:=
\tfrac{1}{T}Y_0\Psi^\top, \nonumber \\ &
\bar Y_1
:=
\tfrac{1}{T}Y_1\Psi^\top,\,
\bar\Phi
:=
\tfrac{1}{T}\Phi\Psi^\top,
\label{eq:normalized_data}
\end{align}
{and, for each $i$, $\bar\Phi^{(i)}:=\tfrac{1}{T}\Phi\Psi^{(i)\top}$ and $\theta_{\rm IV}^{(i)}:=\tfrac{1}{T}Y_1\Psi^{(i)\top}(\bar\Phi^{(i)})^{-1}$.}

\begin{assumption}
\label{ass:PE}
\(
\rank(\bar\Phi^{{(i)}})=n+m
\) {for $i=1,\ldots,$}{$N_\psi$}, and
$\rank(\Phi)=n+m$.
\end{assumption}

\begin{assumption}
\label{ass:exogenous}
The presample and $x_0$ are independent of
$\{\omega_k,\upsilon_k\}_{k\ge0}$, and there is an exogenous excitation
$\{e_k\}$, independent of these and of all noises, such that $u_k$ and {every $\psi_k^{(i)}$ are}
measurable with respect to
\(
\mathcal F_k:=\sigma(y_{{-L}:k-1},u_{{-L}:k-1},e_{{-L}:k}).
\)
\end{assumption}

Under Assumptions~\ref{noises} and~\ref{ass:exogenous},
$\E[\omega_k,\upsilon_k,\upsilon_{k+1}\mid\mathcal F_k]=0$ componentwise and
$\E[\widetilde w_k\psi_k^{(i)\top}]=0$.

\begin{remark}
\label{rem:closed-loop-data}
An instrument must be $\mathcal F_k$-measurable, so that it is uncorrelated with $\tilde w_k$, and relevant, so that $\bar\Phi^{(i)}$ is well conditioned. Under feedback $u_k=Ky_k+e_k$, $u_k$ contains $\upsilon_k$, so $[y_{k-1}^\top\;u_k^\top]^\top$ is invalid and $[y_{k-1}^\top\;e_k^\top]^\top$ can be used instead. Under white open-loop excitation, $\mathbb E[y_ky_{k-1}^\top]=A^\star\mathrm{Cov}(x_{k-1})$, so $\sigma_{\min}(\bar\Phi)$ degrades for fast or near-deadbeat modes. For $m=1$, the past-input instrument {$\psi_k^{(2)}$} gives $\mathbb E[\phi_k\psi_k^{{(2)}\top}]=\sigma_u^2\,\mathrm{diag}([B^\star\;A^\star B^\star\cdots(A^\star)^{n-1}B^\star],1)$, nonsingular whenever $(A^\star,B^\star)$ is controllable. {In open loop it remains valid even when the noise is temporally correlated, whereas $y_{k-1}$ does not.}
\end{remark}


\section{Closed-Loop Parametrization and Uncertainty Quantification}
\subsection{IV-Based Closed-Loop Parametrization}

{The first step writes the closed loop directly in terms of data. The only unknown left in it is the noise that has already occurred along the collected trajectory, which the rest of this section bounds.}

\begin{lemma}
\label{lem:cl_data_param}
Let Assumption~\ref{ass:PE} hold, and let $G_K$ satisfy
\begin{align}
\begin{bmatrix}
I_n\\
K
\end{bmatrix}
=
\bar \Phi G_K,
\label{eq:G_param}
\end{align}
so that $K=\bar U_0G_K$ and $\bar Y_0G_K=I_n$. Then, the closed-loop dynamics of \eqref{eq:effective_disturbance} under $u_k=Ky_k$ are
\begin{align}
y_{k+1}
=
(\bar Y_1-\bar N)G_K\,y_k
+
\widetilde w_k,
\label{eq:cl_state_ABK_tc}
\end{align}
where
\begin{align} \label{eq:normalized_noise}
\bar N:=\frac{1}{T}\widetilde W\Psi^\top, \quad \widetilde W :=\begin{bmatrix}
      \tilde  w_0 & \cdots & \tilde  w_{T-1}
    \end{bmatrix}
\end{align}
collects the effective-noise realizations along the collected trajectory. In particular,
\begin{align}
A^\star+B^\star K
=
(\bar Y_1-\bar N)G_K.
\label{eq:cl_matrix_param}
\end{align}
\end{lemma}

\begin{proof}
Using \eqref{eq:effective_disturbance} and
stacking the collected data in \eqref{eq:collected_data} gives
\(
Y_1
=
A^\star Y_0+B^\star U_0+\widetilde W.
\)
Right-multiplying it by
$\Psi^\top/T$ and using \eqref{eq:normalized_data} and the definition of $\bar N$ in \eqref{eq:normalized_noise} yields
\(
\bar Y_1
=
A^\star\bar Y_0
+
B^\star\bar U_0
+
\bar N.
\)
Hence,
\(
\bar Y_1-\bar N
=
A^\star\bar Y_0+B^\star\bar U_0.
\)
Right-multiplying this equation by $G_K$ and using
\(
\bar Y_0G_K=I
\)
and
\(
\bar U_0G_K=K
\)
gives
 \eqref{eq:cl_matrix_param}. Finally, under the feedback law
\(
u_k=Ky_k
\),
\eqref{eq:effective_disturbance} becomes
\(
y_{k+1}
=
(A^\star+B^\star K)y_k+\widetilde w_k.
\)
Substituting \eqref{eq:cl_matrix_param} into this equation gives
\eqref{eq:cl_state_ABK_tc}. Assumption \ref{ass:PE} guarantees the existence and uniqueness of $G_K$ for every $K$.
\end{proof}

The closed loop \eqref{eq:cl_state_ABK_tc} thus involves the \emph{realized noise} $\bar N$, which has already occurred and can be refined using the trajectory and the prior, and the \emph{aleatory noise} $\widetilde w_k$, whose covariance is what must be learned.

\begin{remark}
\label{rem:one-uncertainty}
With $\theta_{\rm IV}:=\bar Y_1\bar\Phi^{-1}$, the stacked data equation gives
$[\Delta_A\;\Delta_B]:=\theta_{\rm IV}-\theta^\star=\bar N\bar\Phi^{-1}$, so the closed-loop error is exactly
$\Delta_A+\Delta_BK=\bar NG_K$. Independent bounds would give $\|\Delta_A\|_2+\|\Delta_B\|_2\|K\|_2$, treating both worst cases as simultaneous. The model coordinate $\theta$ ties both errors to the same $\bar N$.
\end{remark}

\subsection{{Initial Covariance and Self-Normalized Noise Certificates}}

{Every certificate below scales with the unknown covariance $\Sigma_{\rm true}$ of the effective noise. The next result relates it to the empirical second moment of the realized noise through quantiles that depend only on $n$ and $T$. Since the effective noise is one-dependent, it splits the samples into two independent subsequences.}

\begin{lemma}
\label{lem:cov-lower}
Let Assumption~\ref{noises} hold, let
\(
{\cal I}_q:=\{k\in\{0,\ldots,T-1\}:k\equiv q\pmod 2\}
\) and $T_q:=|\mathcal I_q|$, $q\in\{0,1\}$. For $t\ge n$, let
$G_t\in\R^{n\times t}$ have i.i.d.\ ${\cal N}(0,1)$ entries, and for $p\in(0,1)$ define
\begin{align}
\ell_{n,t}(p)
&:=
\sup\Big\{\ell\ge0:\;
\Prob\big[\lammin\big(\tfrac1tG_tG_t^\top\big)<\ell\big]\le p\Big\},
\nonumber\\
u_{n,t}(p)
&:=
\inf\Big\{u\ge0:\;
\Prob\big[\lammax\big(\tfrac1tG_tG_t^\top\big)>u\big]\le p\Big\}.
\label{eq:pivotal-quantiles}
\end{align}
For $\delta_c\in(0,1)$ and $\lfloor T/2\rfloor\ge n$, let
\begin{align}
\alpha_T:=\sum_{q=0}^1\frac{T_q}{T}\,\ell_{n,T_q}\big(\tfrac{\delta_c}{4}\big),
\quad
\beta_T:=\sum_{q=0}^1\frac{T_q}{T}\,u_{n,T_q}\big(\tfrac{\delta_c}{4}\big).
\label{eq:alpha-beta-exact}
\end{align}
Then, $\alpha_T>0$, $\alpha_T,\beta_T\to1$, and, with probability at least \(1-\delta_c\),
\begin{align}
\alpha_T\Sigma_{\rm true}
\preceq
\frac1T\widetilde W\widetilde W^\top
\preceq
\beta_T\Sigma_{\rm true}.
\label{eq:cov-lower-bound}
\end{align}
\end{lemma}

\begin{proof}
 By
\eqref{effective1}, \(\widetilde w_k\) is a linear function of
\((\omega_k,\upsilon_k,\upsilon_{k+1})\) only. For \(|k-\ell|\ge2\), the
triples \((\omega_k,\upsilon_k,\upsilon_{k+1})\) and
\((\omega_\ell,\upsilon_\ell,\upsilon_{\ell+1})\) share no variable. Any two
distinct indices in the same parity class \({\cal I}_q\) differ by at least
two, so the samples \(\{\widetilde w_k\}_{k\in{\cal I}_q}\) are functions of
disjoint blocks of mutually independent Gaussian variables. Hence, they are
mutually independent and jointly Gaussian, with
\(\E[\widetilde w_k]=0\) and
\(\E[\widetilde w_k\widetilde w_k^\top]=\Sigma_{\rm true}\) by
\eqref{sigmatrue}.

Let
\(\widetilde W_q:=[\widetilde w_k]_{k\in{\cal I}_q}\in\R^{n\times T_q}\) and
\(G_q:=\Sigma_{\rm true}^{-1/2}\widetilde W_q\), which is well defined since
\(\Sigma_{\rm true}\succ0\). By the previous step, the columns of \(G_q\) are
independent \({\cal N}(0,I_n)\) vectors, so \(G_q\) has i.i.d.\
\({\cal N}(0,1)\) entries and
\(\widetilde W_q=\Sigma_{\rm true}^{1/2}G_q\). Thus, \(G_q\) has the same
distribution as \(G_t\) in \eqref{eq:pivotal-quantiles} with \(t=T_q\), and
this distribution does not depend on \(\Sigma_{\rm true}\).

Let
\(
X_q:=\lammin(T_q^{-1}G_qG_q^\top)
\)
and
\(
F(\ell):=\Prob[X_q<\ell].
\)
Since \(T_q\ge n\) and \(G_q\in\mathbb R^{n\times T_q}\) has i.i.d.\
standard Gaussian entries, \(G_q\) has full row rank almost surely.
Hence, \(G_qG_q^\top\succ0\) and therefore \(X_q>0\) almost surely.
It follows that \(F(\ell)\to0\) as \(\ell\downarrow0\), so for some
\(\ell_0>0\),
\(
F(\ell_0)\le\delta_c/4.
\)
Thus, by the definition of the supremum in
\eqref{eq:pivotal-quantiles},
\(
\ell_{n,T_q}(\delta_c/4)\ge\ell_0>0.
\)
Moreover, \(F(\ell)\to1\) as \(\ell\to\infty\), so this supremum is
finite. Hence \(\alpha_T>0\) by \eqref{eq:alpha-beta-exact}.

It remains to verify that the probability bound also holds at the
supremum itself. Since
\(
F(\ell)=\Prob[X_q<\ell]
\)
is nondecreasing and left-continuous, if
\(
\ell_k\uparrow\ell_{n,T_q}(\delta_c/4)
\)
through admissible values, then
\(
F(\ell_k)\le\delta_c/4\) implies
\(
F\!\left(\ell_{n,T_q}(\delta_c/4)\right)\le\delta_c/4.
\)
Therefore,
\[
\Prob\!\left[
\lammin(T_q^{-1}G_qG_q^\top)
<
\ell_{n,T_q}(\delta_c/4)
\right]
\le\delta_c/4.
\]

Likewise,
\(
u\mapsto
\Prob[\lammax(T_q^{-1}G_qG_q^\top)>u]
\)
is nonincreasing and right-continuous. Taking a decreasing sequence of
admissible values approaching the infimum in
\eqref{eq:pivotal-quantiles} gives
\[
\Prob\!\left[
\lammax(T_q^{-1}G_qG_q^\top)
>
u_{n,T_q}(\delta_c/4)
\right]
\le\delta_c/4.
\]
A union bound over the four
events (\(q=0,1\), lower and upper) shows that, with probability at least
\(1-\delta_c\),
\begin{align*}
\ell_{n,T_q}(\delta_c/4)\,I_n\preceq \tfrac1{T_q}G_qG_q^\top\preceq
u_{n,T_q}(\delta_c/4)\,I_n,\qquad q=0,1 .
\end{align*}
Multiplying on the left and on the right by \(\Sigma_{\rm true}^{1/2}\)
preserves the ordering and, since
\(\widetilde W_q=\Sigma_{\rm true}^{1/2}G_q\), gives
\begin{align*}
\ell_{n,T_q}(\delta_c/4)\,\Sigma_{\rm true}
\preceq\tfrac1{T_q}\widetilde W_q\widetilde W_q^\top
\preceq u_{n,T_q}(\delta_c/4)\,\Sigma_{\rm true}.
\end{align*}

The classes \({\cal I}_0\) and
\({\cal I}_1\) partition \(\{0,\ldots,T-1\}\), so
\(\widetilde W\widetilde W^\top
=\widetilde W_0\widetilde W_0^\top+\widetilde W_1\widetilde W_1^\top\) and
\(
\frac1T\widetilde W\widetilde W^\top
=\sum_{q=0}^1\frac{T_q}{T}\Big(\frac1{T_q}\widetilde W_q\widetilde W_q^\top\Big).
\)
Using this equation and summing the bounds above with the nonnegative weights \(T_q/T\) and
using \eqref{eq:alpha-beta-exact} yields \eqref{eq:cov-lower-bound}. 

Finally, by the law of large numbers, \(t^{-1}G_tG_t^\top\to I_n\) in probability as
\(t\to\infty\), so both \(\lammin(t^{-1}G_tG_t^\top)\) and
\(\lammax(t^{-1}G_tG_t^\top)\) converge to $1$ in probability. Their
quantiles therefore also converge to $1$, i.e.,
\(\ell_{n,t}(p)\to1\) and \(u_{n,t}(p)\to1\) for every fixed
\(p\in(0,1)\). Since \(T_q\to\infty\) as \(T\to\infty\) and the weights
\(T_q/T\) sum to one, \(\alpha_T\to1\) and \(\beta_T\to1\).
\end{proof}

\begin{remark}
\label{rem:pivotal-quantiles}
The quantiles in \eqref{eq:pivotal-quantiles} depend only on $n$, $t$ and $p$. They follow from the closed-form extreme-eigenvalue distributions of Wishart matrices~\cite{Chiani2017} by solving one scalar equation each. All results remain valid with looser bounds, e.g.\ Monte Carlo quantiles with an order-statistic margin.
\end{remark}

{The realized noise is unknown, but the prior confines it: every model in $\mathcal M_{\theta,p}$ implies a residual, and the spread of these residuals bounds the realized noise and hence $\Sigma_{\rm true}$.}

\begin{lemma}
\label{lem:init-cert-prior}
Let Assumptions~\ref{noises}--\ref{ass:CMZ-prior} hold, let \(\alpha_T\) be as in Lemma~\ref{lem:cov-lower}, and define
\begin{align}
& C_{\widetilde W,p}
:=
Y_1-C_{\theta,p}\Phi, \quad
\widehat\Sigma_p
:=
\frac1TC_{\widetilde W,p}C_{\widetilde W,p}^\top,
\label{eq:prior-residual}\\
& \mu_p
:=
\frac{1}{\sqrt T}
\sum_{i=1}^{q_{\theta,p}}
\|G_{\theta,p,i}\Phi\|_2, \quad
\varepsilon_0^\star
:=
\frac{\mu_p}{\sqrt{\lammax(\widehat\Sigma_p)}}
.
\label{eq:mu-p}
\end{align}
Then, with probability at least
\(1-\delta_c\),
\begin{align}
\Sigma_{\rm true}
\preceq
\Sigma^{(0)}
:=
\frac{1}{\alpha_T}
\left[
(1+\varepsilon_0^\star)\widehat\Sigma_p
+
(1+{\varepsilon_0^\star}^{-1})\mu_p^2I_n
\right].
\label{eq:init-cert-prior}
\end{align}
\end{lemma}

\begin{proof}
By Assumption~\ref{ass:CMZ-prior}, the true model belongs to the
prior matrix zonotope. Hence, there exists a coefficient vector
\(\zeta^\star\) satisfying
\(\|\zeta^\star\|_\infty\leq1\) such that
\(
\theta^\star
=
C_{\theta,p}
+
\sum_i G_{\theta,p,i}\zeta_i^\star 
\).
Substituting this representation into the stacked data equation
\(
Y_1=\theta^\star\Phi+\widetilde W
\), obtained in Lemma \ref{lem:cl_data_param},
gives 
\begin{align} \label{wide1}
  \widetilde W
=
C_{\widetilde W,p}
-
\sum_iG_{\theta,p,i}\Phi\,\zeta_i^\star ,  
\end{align}
where
\(
C_{\widetilde W,p}
\) is defined in \eqref{eq:prior-residual}.
{Since \(|\zeta_i^\star|\leq1\), the triangle inequality and the
definition of \(\mu_p\) in \eqref{eq:mu-p} give
\(\|\sum_iG_{\theta,p,i}\Phi\,\zeta_i^\star\|_2
\leq\sum_i\|G_{\theta,p,i}\Phi\|_2\leq\sqrt{T}\,\mu_p\).}
Consequently, if
\(
E_p:=
-\sum_iG_{\theta,p,i}\Phi\,\zeta_i^\star,
\)
then
\begin{align}
E_pE_p^\top
\preceq
\|E_p\|_2^2I_n
\preceq
T\mu_p^2I_n, \quad \widetilde W=C_{\widetilde W,p}+E_p .
\label{eq:init-Ep-bound}
\end{align}
where the equality comes from \eqref{wide1}. On the other hand, for any \(\varepsilon_0>0\), the matrix Young inequality gives
\(
\widetilde W\widetilde W^\top
\preceq{}
(1+\varepsilon_0)
C_{\widetilde W,p}C_{\widetilde W,p}^\top
+
(1+\varepsilon_0^{-1})E_pE_p^\top .
\)
Dividing by \(T\), using
\(
\widehat\Sigma_p
\) in \eqref{eq:prior-residual},
and applying \eqref{eq:init-Ep-bound} for its second term yields
\(
\frac1T\widetilde W\widetilde W^\top
\preceq
(1+\varepsilon_0)\widehat\Sigma_p
+
(1+\varepsilon_0^{-1})\mu_p^2I_n .
\)
{Combining this inequality with the lower bound
\(\alpha_T\Sigma_{\rm true}\preceq T^{-1}\widetilde W\widetilde W^\top\)
of Lemma~\ref{lem:cov-lower}, valid with probability at least
\(1-\delta_c\), and dividing by \(\alpha_T>0\) gives
\(\Sigma_{\rm true}\preceq\Sigma^{(0)}:=\alpha_T^{-1}
[(1+\varepsilon_0)\widehat\Sigma_p+(1+\varepsilon_0^{-1})\mu_p^2I_n]\).
It remains to {choose} \(\varepsilon_0\). Since
\(
{\varepsilon_0\widehat\Sigma_p}+\frac{\mu_p^2}{\varepsilon_0}I_n
\preceq
\big({\varepsilon_0}\lambda_p+\frac{\mu_p^2}{\varepsilon_0}\big)I_n,
\,
\lambda_p:=\lambda_{\max}({\widehat\Sigma_p}),
\)
it suffices to minimize the scalar quantity
\(
g(\varepsilon_0)
=
\varepsilon_0\lambda_p+\frac{\mu_p^2}{\varepsilon_0}.
\)
Its unique minimizer is
\(
\varepsilon_0^\star=\frac{\mu_p}{\sqrt{\lambda_p}},
\)
{as in \eqref{eq:mu-p},} which, upon substitution, gives \eqref{eq:init-cert-prior}.}
\end{proof}

{The conservative initial certificate $\Sigma^{(0)}$ is refined in the subsequent sections. The next lemma bounds the projections of the realized, temporally correlated noise onto the delayed instruments, uniformly over all valid covariance certificates.}

\begin{lemma}
\label{lem:iv}
Let Assumptions~\ref{noises}, \ref{ass:PE} and~\ref{ass:exogenous} hold, let $d_z:=n+m$, fix $\lambda_\psi>0$ and {$\delta_{s,i}>0$ with $\sum_{i=1}^{{N_\psi}}\delta_{s,i}=\delta_s$}, and define, for each instrument,
\begin{align}
V_\psi{^{(i)}}
&:=
\lambda_\psi I_{d_z}
+
\sum_{k=0}^{T-1}\psi_k {^{(i)}}\psi_k {^{(i)\top}},
\nonumber\\
\gamma(\delta)
&:=
\Big[
n\log
\frac{\det(V_\psi {^{(i)}})}{\lambda_\psi^{d_z}}
+
2\log\frac2{\delta}
\Big]^{1/2},
\nonumber\\
D_\psi {^{(i)}}
&:=
\frac{\sqrt2\,\gamma( {\delta_{s,i}})}{T}(V_\psi {^{(i)}})^{1/2},
\label{eq:Dpsi}
\end{align}
{with $V_\psi:=V_\psi^{(1)}$ and $D_\psi:=D_\psi^{(1)}$.} Then, with probability at least \(1-\delta_s\), {for every $i$ and} every $\Sigma^{(j)}$ with $\Sigma_{\rm true}\preceq\Sigma^{(j)}$,
\begin{align}
& {T^{-1}\widetilde W\Psi^{(i)\top}}
=
C_j\Delta_j{^{(i)}}D_\psi {^{(i)}},
\nonumber \\ &
\|\Delta_j{^{(i)}}\|_2\leq1,
\,
C_j:=(\Sigma^{(j)})^{1/2}.
\label{eq:selfnormalized-j}
\end{align}
In particular, $\bar N=C_j\Delta_jD_\psi$ for $i=1$.
\end{lemma}

\begin{proof}
{Decompose the effective noise as}
\(
\widetilde w_k={\nu_k}+\upsilon_{k+1},
\,
{\nu_k}:=\omega_k-A^\star\upsilon_k,
\)
where
\(
{\Sigma_\nu}:=W+A^\star V(A^\star)^\top
\) is the covariance of {$\nu_k$}.
Then,
\(
\Sigma_{\mathrm{true}}={\Sigma_\nu}+V.
\)
Let {$\Pi_\nu$} and $\Pi_V$ be the orthogonal
projectors onto the ranges of {$\Sigma_\nu$} and $V$, let $\xi_k,\xi'_k\sim\mathcal N(0,I_n)$
be auxiliary i.i.d.\ sequences independent of all other variables, and define
\(
g_k:=({\Sigma_\nu}^{1/2})^\dagger {\nu_k}+(I_n-{\Pi_\nu})\xi_k,
\,
h_k:=(V^{1/2})^\dagger\upsilon_{k+1}+(I_n-\Pi_V)\xi'_k .
\)
Then, $g_k\sim\mathcal N(0,I_n)$, $h_k\sim\mathcal N(0,I_n)$, and, almost surely,
{$\nu_k=\Sigma_\nu^{1/2}g_k$} and $\upsilon_{k+1}=V^{1/2}h_k$.

{
{Let $\{\mathcal G_k\}$ be a filtration such that $\psi_k$ is
$\mathcal G_k$-measurable and $q_k\in\mathbb R^n$ is
$\mathcal G_{k+1}$-measurable, $\mathcal N(0,I_n)$, and independent of
$\mathcal G_k$. Then, $q_k \, \psi^\top$ is a martingale-difference-type matrix increment. Thus, by the standard self-normalized martingale
argument based on the method of mixtures
\cite{abbasi2011online}, for
\(
S_T:=\sum_{k=0}^{T-1}q_k\psi_k^\top
\)
and $V_\psi$ in \eqref{eq:Dpsi},
\begin{align}
\Prob\!\left[
\|S_TV_\psi^{-1/2}\|_F^2
>
n\log\!\frac{\det V_\psi}{\lambda_\psi^{d_z}}
+
2\log\!\frac1\delta
\right]
\leq\delta .
\label{eq:matrix-self-normalized}
\end{align}
The result follows by applying the exponential-supermartingale
construction row-wise and combining the rows through the Gaussian
mixture argument.}

{The filtration $\{\mathcal F_k\}$ of Assumption \ref{ass:exogenous} cannot serve as $\{\mathcal G_k\}$,
because $\omega_k$ and $\upsilon_k$ are not $\mathcal F_{k+1}$-measurable. Let
$\mathcal H$ collect $x_0$ and the presample history, and let
$\mathcal G_k^{g}:=\sigma(\mathcal H,e_{0:k},\omega_{0:k-1},\upsilon_{0:k-1},\xi_{0:k-1})$,
$\mathcal G_k^{h}:=\sigma(\mathcal H,e_{0:k},\omega_{0:k},\upsilon_{0:k},\xi'_{0:k-1})$.
Both contain $\mathcal F_k$, so every admissible $\psi_k^{(i)}$ is measurable with respect to each. By
Assumptions~\ref{noises} and~\ref{ass:exogenous}, $g_k$, a function of
$(\omega_k,\upsilon_k,\xi_k)$, is $\mathcal G_{k+1}^{g}$-measurable and
independent of $\mathcal G_k^{g}$, and $h_k$, a function of
$(\upsilon_{k+1},\xi'_k)$, is $\mathcal G_{k+1}^{h}$-measurable and independent of
$\mathcal G_k^{h}$.} Fix an instrument $i$ and drop its index, so that $\psi_k$, $V_\psi$ and $\delta_s$ stand for $\psi_k^{(i)}$, $V_\psi^{(i)}$ and $\delta_{s,i}$. Applying
\eqref{eq:matrix-self-normalized} to each
sequence {with its own filtration} at level $\delta_{s,i}/2$, using $\|X\|_2\le\|X\|_F$, and taking a union
bound gives, with probability at least $1-\delta_s$, matrices
$\Delta_1,\Delta_2$ with $\|\Delta_i\|_2\le1$ such that
\[
\sum_{k=0}^{T-1}g_k\psi_k^\top
=\gamma(\delta_s)\Delta_1V_\psi^{1/2},\quad
\sum_{k=0}^{T-1}h_k\psi_k^\top
=\gamma(\delta_s)\Delta_2V_\psi^{1/2}.
\]
}
{Multiplying these on the left by {$\Sigma_\nu^{1/2}$} and $V^{1/2}$,
respectively, and adding gives}
\[
\bar N
=
\frac{1}{T}\sum_{k=0}^{T-1}\widetilde w_k\psi_k^\top
=
\frac{\gamma(\delta_s)}{T}
\left(
{\Sigma_\nu^{1/2}}\Delta_1
+
V^{1/2}\Delta_2
\right)V_\psi^{1/2}.
\]
Define
\(
{\Lambda_1}:=\Sigma_{\mathrm{true}}^{-1/2}{\Sigma_\nu^{1/2}},
\,
{\Lambda_2}:=\Sigma_{\mathrm{true}}^{-1/2}V^{1/2}
\){, which are well defined since $\Sigma_{\rm true}\succ0$}.
Since {$\Sigma_\nu$}$+V=\Sigma_{\mathrm{true}}$, one has
\(
{\Lambda_1\Lambda_1^\top+\Lambda_2\Lambda_2^\top}=I_n,
\)
and hence
\(
\left\|
\begin{bmatrix}
{\Lambda_1} & {\Lambda_2}
\end{bmatrix}
\right\|_2=1.
\)
Now, define
\[
\Delta^\star
:=
\frac{1}{\sqrt{2}}
\begin{bmatrix}
{\Lambda_1} & {\Lambda_2}
\end{bmatrix}
\begin{bmatrix}
{\Delta_1}\\
{\Delta_2}
\end{bmatrix}, \quad \left\|
\begin{bmatrix}
{\Delta_1}\\
{\Delta_2}
\end{bmatrix}
\right\|_2
\le \sqrt{2},
\]
where the inequality is obtained from $\|\Delta_i\|_2\le1$.
Therefore,
\(
\|\Delta^\star\|_2\le1.
\)
{Since {$\Sigma_\nu^{1/2}$}$\Delta_1+V^{1/2}\Delta_2
=\Sigma_{\rm true}^{1/2}({\Lambda_1}\Delta_1+{\Lambda_2}\Delta_2)$,} one has
\[
\bar N
=
\Sigma_{\mathrm{true}}^{1/2}
\Delta^\star
\frac{\sqrt{2}\,\gamma(\delta_s)}{T}
V_\psi^{1/2}
=
\Sigma_{\mathrm{true}}^{1/2}\Delta^\star D_\psi.
\]
Since
\(
\Sigma_{\mathrm{true}}\preceq\Sigma^{(j)},
\)
there exists a contraction matrix {$\Lambda$} with
{$\|\Lambda\|_2\le1$} such that
\(
\Sigma_{\mathrm{true}}^{1/2}
=
C_j{\Lambda}
\), a pathwise step valid for all such $\Sigma^{(j)}$, so
\(
T^{-1}\widetilde W\Psi^\top=C_j(\Lambda\Delta^\star)D_\psi
\) with
\(
\|{\Lambda}\Delta^\star\|_2\le1.
\)
A union bound over $i$ completes the proof.
\end{proof}

{Since the projected noise is an affine function of the model, this bound transfers to a confidence set for $\theta$, one per instrument.}

\begin{lemma}
\label{lem:selfnormalized-MZ}
On the confidence event of Lemma~\ref{lem:iv}, with
\begin{align} \label{thetaIV}
\theta_{\rm IV}:=\theta_{\rm IV}^{(1)}=\bar Y_1\bar\Phi^{-1},
\end{align}
the true model belongs to {$\bigcap_{i=1}^{{N_\psi}}\Theta_{\rm sn}^{(i,j)}$, where}
\begin{align}
\Theta_{\rm sn}^{ {(i,j)}}
:=
\Big\{
\theta:\;
\big\|C_j^{-1}(\theta_{\rm IV} {^{(i)}}-\theta)\bar\Phi {^{(i)}} (D_\psi {^{(i)}})^{-1}\big\|_2\le1
\Big\}
\label{eq:theta-SN-MZ}
\end{align}
is convex and compact. {We write $\Theta_{\rm sn}^{(j)}:=\Theta_{\rm sn}^{(1,j)}$.}
\end{lemma}

\begin{proof}
By Lemma~\ref{lem:cl_data_param}, $T^{-1}Y_1\Psi^{(i)\top}=\theta^\star\bar\Phi^{(i)}+T^{-1}\widetilde W\Psi^{(i)\top}$, so
$T^{-1}\widetilde W\Psi^{(i)\top}=(\theta_{\rm IV}^{(i)}-\theta^\star)\bar\Phi^{(i)}$, and \eqref{eq:selfnormalized-j} gives the claim. The set is the preimage of the unit spectral ball under an affine map, and $\|\theta_{\rm IV}^{(i)}-\theta\|_2\le\|C_j\|_2\|D_\psi^{(i)}\|_2\|(\bar\Phi^{(i)})^{-1}\|_2$ on it.
\end{proof}

{Instrument~1 centers the cover below and enters all rate statements. The others only remove models. Since $\gamma(\delta)$ grows like $\sqrt{\log(1/\delta)}$, giving most of $\delta_s$ to instrument~1 keeps $\Theta_{\rm sn}^{(j)}$ almost unchanged. With} {$N_\psi\ge2$, the $N_\psi n(n+m)$} {moment conditions overidentify the $n(n+m)$ model parameters.}

\begin{remark}
\label{rem:LS-bias}
The least-squares estimate
\begin{align}
\theta_{\rm LS}
:=
Y_1\Phi^\dagger,
\qquad
Q_T:=
\tfrac{1}{T}\Phi\Phi^\top,
\label{thetaLs}
\end{align}
satisfies $\theta_{\rm LS}-\theta^\star=T^{-1}\widetilde W\Phi^\top Q_T^{-1}\to C_{w\phi}Q^{-1}$ when $Q_T\to Q\succ0$, with $C_{w\phi}:=\mathbb E[\widetilde w_k\phi_k^\top]\neq0$ in general, so its bias does not vanish with more data, and the residual covariance $T^{-1}\sum_kr_kr_k^\top$, $r_k=\widetilde w_k+(\theta^\star-\theta)\phi_k$, contains indefinite cross terms and need not dominate $\Sigma_{\rm true}$. The delayed instruments satisfy $\E[\widetilde w_k\psi_k^{(i)\top}]=0$, which is what makes Lemmas~\ref{lem:iv}--\ref{lem:selfnormalized-MZ} valid.
\end{remark}

\noindent Let $\mathcal E_{\rm cov}$ and $\mathcal E_{\rm sn}$ denote the
confidence events of Lemmas~\ref{lem:cov-lower} and~\ref{lem:iv}, and
$\mathcal E_{\rm all}:=\mathcal E_{\rm cov}\cap
\mathcal E_{\rm sn}$, with
$\Prob(\mathcal E_{\rm all})\ge1-(\delta_c+\delta_s)$.
Since these bounds hold for all valid covariance certificates, refinement on the same data needs no further confidence allocation.

\subsection{Data- and Prior-Consistent Refinement}

For a candidate model $\theta$, the data equation
$Y_1=\theta^\star\Phi+\widetilde W^\star$ motivates the residual map
\begin{align}
\widetilde W(\theta):=Y_1-\theta\Phi,
\label{eq:residual-map}
\end{align}
which returns the unique realized-noise sequence that $\theta$
requires to reproduce the trajectory. $\widetilde W(\theta^\star)=\widetilde W^\star$. The data- and prior-consistent realized-noise sequences are
\begin{align}
\widetilde{\mathcal W}_{dp}
:=
\left\{
Y_1-\theta\Phi:
\theta\in\mathcal M_{\theta,p}
\right\}\ni\widetilde W^\star.
\label{eq:data-prior-noise-set}
\end{align}
Since $\Phi$ has full row rank, \eqref{eq:residual-map} is injective, so $\widetilde{\mathcal W}_{dp}$ is in one-to-one correspondence with $\mathcal M_{\theta,p}$ and lies in an affine subspace of dimension $n(n+m)$, independent of $T$: each sample adds $n$ noise entries but also $n$ equations tying them to the same model. {Trajectory consistency is therefore an equality constraint: a noise matrix $\widetilde W$ reproduces the data with some model if and only if $(Y_1-\widetilde W)\Phi^\perp=0$, which is $nT-n(n+m)$ linear equalities, and the $\theta$-coordinates eliminate it exactly by parametrizing all of its solutions. It removes no model on its own. Refinement comes from the prior and from statistical moment conditions on $\widetilde W(\theta)$.}

{The moment conditions are those of Lemma~\ref{lem:iv}: with $C_jC_j^\top=\Sigma^{(j)}$, the delayed-IV components of the true realization satisfy a self-normalized bound for every instrument.  Therefore, similar to \eqref{eq:theta-SN-MZ},  define the statistically admissible realizations}
\begin{align}
& \widetilde{\mathcal W}_{s}^{(j)}
:= \nonumber \\ &
\Big\{
\widetilde W:\;
\big\|
C_j^{-1}
\big(\tfrac1T\widetilde W {\Psi^{(i)\top}}\big)
 {(D_\psi^{(i)})^{-1}}
\big\|_2\leq1,\ {i=1,\ldots,}{N_\psi}
\Big\},
\label{eq:statistical-noise-set}
\end{align}
which contains $\widetilde W^\star$ on $\mathcal E_{\rm all}$ whenever $\Sigma_{\rm true}\preceq\Sigma^{(j)}$, and the refined set
\begin{align}
\widetilde{\mathcal W}_{r}^{(j)}
:=
\widetilde{\mathcal W}_{dp}
\cap
\widetilde{\mathcal W}_{s}^{(j)}.
\label{eq:refined-noise-set}
\end{align}

{The following identities express the noise of a model, its projections and its second moment in the model coordinate, so that no noise variable is needed.}

\begin{lemma}
\label{lem:residual-identity}
Let Assumption~\ref{ass:PE} hold, and let
$\widehat\Sigma_{\rm LS}
:=
T^{-1}
\widetilde W(\theta_{\rm LS})
\widetilde W(\theta_{\rm LS})^\top$. Then:
\begin{enumerate}
\item[(i)] $\widetilde W\in\mathbb R^{n\times T}$ satisfies
$(Y_1-\widetilde W)\Phi^\perp=0$ if and only if
$\widetilde W=\widetilde W(\theta)$ for the unique
$\theta=(Y_1-\widetilde W)\Phi^\dagger$.
\item[(ii)] for every $\theta\in\mathbb R^{n\times(n+m)}$,
\begin{align}
\tfrac1T
\widetilde W(\theta)\widetilde W(\theta)^\top
=
\widehat\Sigma_{\rm LS}
+
(\theta_{\rm LS}-\theta)
Q_T
(\theta_{\rm LS}-\theta)^\top.
\label{eq:residual-identity}
\end{align}
\item[(iii)] $N(\theta):=\tfrac1T\widetilde W(\theta)\Psi^\top
=(\theta_{\rm IV}-\theta)\bar\Phi$ {, and
$\tfrac1T\widetilde W(\theta)\Psi^{(i)\top}=(\theta_{\rm IV}^{(i)}-\theta)\bar\Phi^{(i)}$
for every $i$}.
\end{enumerate}
\end{lemma}

\begin{proof}
We first prove part (i). The condition $(Y_1-\widetilde W)\Phi^\perp=0$ holds if and only if
every row of $Y_1-\widetilde W$ belongs to
$\ker(\Phi)^\perp=\operatorname{row}(\Phi)$. Hence,
$Y_1-\widetilde W=\theta\Phi$ for some $\theta$, which is equivalent to
$\widetilde W=\widetilde W(\theta)$. Since $\Phi$ has full row rank by
Assumption~\ref{ass:PE}, right multiplication by $\Phi^\dagger$ gives
$\theta=(Y_1-\widetilde W)\Phi^\dagger$, establishing uniqueness.
To prove part (ii), use \eqref{eq:residual-map} to write
\(
\widetilde W(\theta)
=
Y_1-\theta\Phi
=
Y_1-\theta_{\rm LS}\Phi
+
(\theta_{\rm LS}-\theta)\Phi
\).
The least-squares normal equation gives
\(
\widetilde W(\theta_{\rm LS})\Phi^\top=0,
\)
and therefore its transpose also gives
\(
\Phi\widetilde W(\theta_{\rm LS})^\top=0.
\)
Hence, after expanding
\(\widetilde W(\theta)\widetilde W(\theta)^\top\), both cross terms vanish.
Dividing by \(T\) and using
\(Q_T=T^{-1}\Phi\Phi^\top\) in \eqref{thetaLs} gives
\eqref{eq:residual-identity}.
Finally, to prove part
(iii), using \eqref{eq:residual-map} and the normalized data matrices gives
$N(\theta)=\bar Y_1-\theta\bar\Phi$. Since
$\theta_{\rm IV}\bar\Phi=\bar Y_1$ by \eqref{thetaIV}, the claim follows.
\end{proof}

Hence $\widetilde{\mathcal W}_{r}^{(j)}$ needs no noise variables: each $\theta\in\mathcal M_{\theta,p}$ generates one $\widetilde W(\theta)\in\widetilde{\mathcal W}_{dp}$, and we only test whether it lies in $\widetilde{\mathcal W}_{s}^{(j)}$. The refined model set is
\begin{align}
\Theta_r^{(j)}
:=
\big\{\theta\in\mathcal M_{\theta,p}:
\widetilde W(\theta)\in\widetilde{\mathcal W}_{s}^{(j)}\big\}.
\label{eq:refined-model-set-conceptual}
\end{align}

\begin{theorem}
\label{thm:reduced-set}
Fix $j$, with
$C_j=(\Sigma^{(j)})^{1/2}\succ0$. Then:
\begin{enumerate}
\item[(i)] the set \eqref{eq:refined-model-set-conceptual} is
\begin{align}
& \Theta_r^{(j)}
= 
\Big\{
\theta\in\mathcal M_{\theta,p}: \nonumber \\ &
\big\|
C_j^{-1}
({\theta_{\rm IV}^{(i)}}-\theta)
{\bar\Phi^{(i)}(D_\psi^{(i)})^{-1}}
\big\|_2\leq1,\ {i=1,\ldots,}{N_\psi}
\Big\}.
\label{eq:reduced-set}
\end{align}
\item[(ii)] $\Theta_r^{(j)}$ is convex and compact, and every $\theta\in\Theta_r^{(j)}$ generates a unique trajectory-consistent noise $\widetilde W(\theta)$.
\item[(iii)] the refined noise and model sets correspond one-to-one. That is,
\begin{align}
\widetilde{\mathcal W}_{r}^{(j)}
=
\big\{
\widetilde W(\theta):
\theta\in\Theta_r^{(j)}
\big\}.
\label{eq:refined-noise-model-equivalence}
\end{align}
\item[(iv)] on $\mathcal E_{\rm all}$, if $\Sigma_{\rm true}\preceq\Sigma^{(j)}$, then
$\theta^\star\in\Theta_r^{(j)}$ and
$\widetilde W(\theta^\star)\in
\widetilde{\mathcal W}_{r}^{(j)}$.
\item[(v)] every $\theta\in\Theta_r^{(j)}$ uniquely determines
$N(\theta)$,
$\widehat\Sigma(\theta):=
T^{-1}\widetilde W(\theta)\widetilde W(\theta)^\top$,
$\theta=[A(\theta)\;B(\theta)]$, and
$A_K(\theta)=A(\theta)+B(\theta)K=(\bar Y_1-N(\theta))G_K$.
\end{enumerate}
\end{theorem}

\begin{proof}
By definition,
$\theta\in\Theta_r^{(j)}$ if and only if $\theta\in\mathcal M_{\theta,p}$
and $\widetilde W(\theta)$ satisfies \eqref{eq:statistical-noise-set}. By
Lemma~\ref{lem:residual-identity}(iii),
$T^{-1}\widetilde W(\theta)\Psi^{{(i)}\top}=(\theta_{\rm IV}^{{(i)}}-\theta)\bar\Phi^{{(i)}}$, so
{each} self-normalized condition in \eqref{eq:statistical-noise-set} is identical to {the corresponding} inequality
in \eqref{eq:reduced-set}, {which proves (i). Each constraint is a norm sublevel set, so their intersection with the compact
matrix zonotope $\mathcal M_{\theta,p}$ is convex and compact.}
The remaining claims in parts (ii) and (iii) follow directly from the injectivity of the
residual map in Lemma~\ref{lem:residual-identity}(i) and from
\eqref{eq:data-prior-noise-set}--\eqref{eq:refined-model-set-conceptual}.

For part (iv), on the confidence event of
Lemma~{\ref{lem:selfnormalized-MZ}}, the true {realization satisfies}
\eqref{eq:statistical-noise-set} {whenever $\Sigma_{\rm true}\preceq\Sigma^{(j)}$}. Since
$\theta^\star\in\mathcal M_{\theta,p}$,
$\widetilde W(\theta^\star)\in
\widetilde{\mathcal W}_{r}^{(j)}$, and
\eqref{eq:refined-noise-model-equivalence} then gives
$\theta^\star\in\Theta_r^{(j)}$.
For part (v), each quantity is an explicit single-valued function of
$\theta$ and the data: by Lemma~\ref{lem:residual-identity}(iii) and (ii),
\(N(\theta)=(\theta_{\rm IV}-\theta)\bar\Phi\) and
\(\widehat\Sigma(\theta)=\widehat\Sigma_{\rm LS}
+(\theta_{\rm LS}-\theta)Q_T(\theta_{\rm LS}-\theta)^\top\),
while the partition gives
\(A(\theta)=\theta\,[I_n\;\,0]^\top\),
\(B(\theta)=\theta\,[0\;\,I_m]^\top\), and
\(A_K(\theta)=\theta\,[I_n\;\,K^\top]^\top
=(\bar Y_1-N(\theta))G_K\) by \eqref{eq:G_param}.
\end{proof}

The refined set splits into a polytopic and a quadratic part,
\begin{align}
\Theta_r^{(j)}=\mathcal M_{\theta,p}\cap\Theta_{\rm s}^{(j)},
\,
\Theta_{\rm s}^{(j)}
:=
\big\{\theta:\widetilde W(\theta)\in\widetilde{\mathcal W}_s^{(j)}\big\},
\label{eq:stat-model-set}
\end{align}
{and only $\Theta_{\rm s}^{(j)}$ is covered, by a constrained matrix zonotope (CMZ). Finitely many generators make the covariance envelope below computable, because the envelope condition is affine in the generator coefficients. The cover starts from the exact interval hull of the primary instrument's set, is tightened by halfspaces that support $\Theta_{\rm s}^{(j)}$, and is intersected with the prior exactly.}

\begin{lemma}
\label{lem:cover}
Let $\{E_s\}_{s=1}^{q_\theta}$, $q_\theta=n(n+m)$,
be an orthonormal basis of $\mathbb R^{n\times(n+m)}$
under the Frobenius inner product, let
\(
M_\psi:=D_\psi\bar\Phi^{-1}
\),
and, for $s=1,\ldots,q_\theta$,
\begin{align}
c_s
:=\langle E_s,\theta_{\rm IV}\rangle,
\quad
w_s
:=\|C_j^\top E_s M_\psi^\top\|_*,
\label{eq:analytic-support}
\end{align}
\begin{align}
C_\theta^{(j)}
:=
\sum_{s=1}^{q_\theta}c_sE_s, \quad
G_{\theta,s}^{(j)}
:=
w_sE_s,
\label{eq:cover-generators}
\end{align}
\begin{align}
\Theta_{\rm z}^{(j)}
:=
\Big\{
C_\theta^{(j)}
+\sum_{s=1}^{q_\theta}G_{\theta,s}^{(j)}\zeta_s:
\|\zeta\|_\infty\leq1
\Big\}.
\label{eq:mz-hull}
\end{align}
For $n_c\geq0$ cut directions
$L_r\in\mathbb R^{n\times(n+m)}$, let
\begin{align}
\overline h_r
{\ \geq\ }
\max_{\theta\in\Theta_{\rm s}^{(j)}}
\langle L_r,\theta\rangle,
\qquad r=1,\ldots,n_c,
\label{eq:cut-support}
\end{align}
{e.g.\ the maximum itself (one SDP) or the closed form of part~(iv)}, and
\begin{align}
&d_r:=\langle L_r,C_\theta^{(j)}\rangle,
\,
[A_c^{(j)}]_{rs}
:=\langle L_r,G_{\theta,s}^{(j)}\rangle,
\,
s_r
:=\sum_{s}
\big|[A_c^{(j)}]_{rs}\big|,
\nonumber\\
&B_c^{(j)}
:=\diag(\eta_r),\
\eta_r
:=\tfrac{s_r+\overline h_r-d_r}{2},
\
[b_c^{(j)}]_r
:=\tfrac{\overline h_r-d_r-s_r}{2},
\label{eq:cmz-data}
\end{align}
\begin{align}
\Theta_{\rm cz}^{(j)}
:=
\Big\{
C_\theta^{(j)}
+\sum_{s=1}^{q_\theta}G_{\theta,s}^{(j)}\zeta_s:\;
&\|\zeta\|_\infty\leq1,\;
\|\xi\|_\infty\leq1,
\nonumber\\
&A_c^{(j)}\zeta+B_c^{(j)}\xi=b_c^{(j)}
\Big\}.
\label{eq:cmz-cover}
\end{align}
Then:
\begin{enumerate}
\item[(i)]
$\Theta_{\rm s}^{(j)}
\subseteq
\Theta_{\rm sn}^{(j)}
\subseteq
\Theta_{\rm z}^{(j)}$, and $\Theta_{\rm z}^{(j)}$ is the exact
interval hull of
$\Theta_{\rm sn}^{(j)}$ in the basis $\{E_s\}$.
\item[(ii)]
If $\Theta_{\mathrm{s}}^{(j)} \neq \emptyset$, $\eta_r\geq0$ for every $r$, and
\begin{align}
\Theta_{\rm cz}^{(j)}
=
\Theta_{\rm z}^{(j)}
\cap
\bigcap_{r=1}^{n_c}
\big\{\theta:
\langle L_r,\theta\rangle\leq\overline h_r
\big\}.
\label{eq:cmz-halfspace}
\end{align}
\item[(iii)]
$\Theta_{\rm s}^{(j)}
\subseteq
\Theta_{\rm cz}^{(j)}
\subseteq
\Theta_{\rm z}^{(j)}$, strictly
if $\overline h_r<d_r+s_r$ for some $r$.
\item[{(iv)}] {with
$\mathcal B(c;P,R):=\{c-P\Delta R:\|\Delta\|_2\le1\}$, whose support is
$\max_{\theta\in\mathcal B}\langle L,\theta\rangle=\langle L,c\rangle+\|P^\top LR^\top\|_*$,
one has $\Theta_{\rm sn}^{(i,j)}=\mathcal B(\theta_{\rm IV}^{(i)};C_j,M_\psi^{(i)})$,
$M_\psi^{(i)}:=D_\psi^{(i)}(\bar\Phi^{(i)})^{-1}$. Hence, the
smallest of these supports {over $i$} in direction $L_r$ is a valid
$\overline h_r$ and needs no SDP.}
\end{enumerate}
\end{lemma}

\begin{proof}
\emph{Part (i).}
By Lemma~\ref{lem:selfnormalized-MZ},
every $\theta\in\Theta_{\rm sn}^{(j)}$
can be represented as
\(
\theta
=
\theta_{\rm IV}-C_j\Delta M_\psi,
\,
\|\Delta\|_2\leq1.
\)
Hence, for any basis direction $E_s$,
\(
\langle E_s,\theta\rangle
=
\langle E_s,\theta_{\rm IV}\rangle
-\langle C_j^\top E_sM_\psi^\top,\Delta\rangle.
\)
The second term is linear in $\Delta$. By duality between the spectral
and nuclear norms,
\(
\max_{\|\Delta\|_2\leq1}
\big|
\langle C_j^\top E_sM_\psi^\top,\Delta\rangle
\big|
=
\|C_j^\top E_sM_\psi^\top\|_*
=
w_s,
\)
and this maximum is attained. Therefore,
\(
\underline g_s
\leq
\langle E_s,\theta\rangle
\leq
\overline g_s,
\,
\forall\theta\in\Theta_{\rm sn}^{(j)},
\)
with both bounds attained over
$\Theta_{\rm sn}^{(j)}$.
Since $\{E_s\}$ is an orthonormal basis,
$\Theta_{\rm z}^{(j)}$ in \eqref{eq:mz-hull}
is exactly the set obtained by allowing each coordinate
$\langle E_s,\theta\rangle$ to vary independently over
$[\underline g_s,\overline g_s]$. Thus,
$\Theta_{\rm z}^{(j)}$ is the coordinatewise interval hull of
$\Theta_{\rm sn}^{(j)}$, and thus $\Theta_{\rm sn}^{(j)}
\subseteq
\Theta_{\rm z}^{(j)}$.
Finally, by the definition of $\Theta_{\rm s}^{(j)}$ in
\eqref{eq:stat-model-set},
\(
\Theta_{\rm s}^{(j)}
\subseteq
\Theta_{\rm sn}^{(j)},
\)
which proves (i).

\emph{Part (ii).}
For any
$\theta(\zeta)=C_\theta^{(j)}
+\sum_sG_{\theta,s}^{(j)}\zeta_s
\in\Theta_{\rm z}^{(j)}$,
the definitions in \eqref{eq:cmz-data} give
\begin{align*}
\langle L_r,\theta(\zeta)\rangle
=d_r+[A_c^{(j)}\zeta]_r,
\quad
-s_r\leq[A_c^{(j)}\zeta]_r\leq s_r.
\end{align*}
Since $\Theta_{\rm s}^{(j)}$ is nonempty and
contained in $\Theta_{\rm z}^{(j)}$, its support
value, {and hence any $\overline h_r$ in \eqref{eq:cut-support},}
satisfies
$\overline h_r\geq d_r-s_r$,
which implies $\eta_r\geq0$.

The supporting halfspace
$\langle L_r,\theta\rangle\leq\overline h_r$
is equivalent, for $\theta=\theta(\zeta)$, to
$[A_c^{(j)}\zeta]_r\leq\overline h_r-d_r$.
Since the lower bound
$[A_c^{(j)}\zeta]_r\geq-s_r$
already holds over the box, imposing this
halfspace amounts to restricting
$[A_c^{(j)}\zeta]_r$ to the interval
$[-s_r,\overline h_r-d_r]$.
By \eqref{eq:cmz-data},
\begin{align*}
[b_c^{(j)}]_r-\eta_r=-s_r,
\qquad
[b_c^{(j)}]_r+\eta_r=\overline h_r-d_r.
\end{align*}
Thus, $[b_c^{(j)}]_r$ and $\eta_r$
are respectively the midpoint and half-width
of this interval. Introducing an auxiliary
variable $\xi_r\in[-1,1]$, the equality
\(
[A_c^{(j)}\zeta]_r+\eta_r\xi_r
=[b_c^{(j)}]_r
\)
is feasible if and only if
$[A_c^{(j)}\zeta]_r$ belongs to this interval,
or equivalently,
$\langle L_r,\theta(\zeta)\rangle
\leq\overline h_r$.
Therefore, the $r$-th equality constraint
in \eqref{eq:cmz-cover} represents exactly
the $r$-th supporting halfspace over
$\Theta_{\rm z}^{(j)}$.
{Since $B_c^{(j)}$ is diagonal, the $n_c$ halfspaces are imposed independently, which proves}
\eqref{eq:cmz-halfspace}.
\emph{Part (iii).}
By part (i),
$\Theta_{\rm s}^{(j)}
\subseteq\Theta_{\rm z}^{(j)}$.
Moreover, by the definition of
$\overline h_r$ in \eqref{eq:cut-support},
every $\theta\in\Theta_{\rm s}^{(j)}$
satisfies
$\langle L_r,\theta\rangle\leq\overline h_r$
for all $r=1,\ldots,n_c$.
Hence, every such $\theta$ also belongs to
the intersection in \eqref{eq:cmz-halfspace},
giving
$\Theta_{\rm s}^{(j)}
\subseteq\Theta_{\rm cz}^{(j)}$.
The inclusion
$\Theta_{\rm cz}^{(j)}
\subseteq\Theta_{\rm z}^{(j)}$
follows directly from
\eqref{eq:cmz-halfspace}.

Finally,
\(
\max_{\theta\in\Theta_{\rm z}^{(j)}}
\langle L_r,\theta\rangle
=d_r+s_r.
\)
If $\overline h_r<d_r+s_r$,
a maximizer over $\Theta_{\rm z}^{(j)}$
violates the $r$-th supporting halfspace
and therefore does not belong to
$\Theta_{\rm cz}^{(j)}$.
This proves strict inclusion.

{\emph{Part (iv).} The form of $\Theta_{\rm sn}^{(i,j)}$ is
Lemma~\ref{lem:selfnormalized-MZ}. The support
follows from
$\langle L,c-P\Delta R\rangle=\langle L,c\rangle-\langle P^\top LR^\top,\Delta\rangle$
and spectral--nuclear duality, and $\Theta_{\rm s}^{(j)}$ lies in each {$\Theta_{\rm sn}^{(i,j)}$}.}
\end{proof}

{The prior is already a matrix zonotope, so it is intersected with the cover exactly.}

\begin{lemma}
\label{lem:full-cover}
With $[c_p]_s:=\langle E_s,C_{\theta,p}\rangle$, $[P]_{si}:=\langle E_s,G_{\theta,p,i}\rangle$ and $c:=\col(c_1,\ldots,c_{q_\theta})$, let
\begin{align}
\bar A_c^{(j)}
:=
\begin{bmatrix}
A_c^{(j)}\\
\diag(w_s)
\end{bmatrix},\
\bar B_c^{(j)}
:=
\begin{bmatrix}
B_c^{(j)}&0\\
0&-P
\end{bmatrix},\
\bar b_c^{(j)}
:=
\begin{bmatrix}
b_c^{(j)}\\
c_p-c
\end{bmatrix},
\label{eq:full-cover-data}
\end{align}
and, with
$\bar\xi:=\operatorname{col}(\xi,\zeta^p)$,
\begin{align}
\Theta_{\rm pcz}^{(j)}
:=
\Big\{
C_\theta^{(j)}
+\sum_{s=1}^{q_\theta}
G_{\theta,s}^{(j)}\zeta_s:\;
&\|\zeta\|_\infty\leq1,\;
\|\bar\xi\|_\infty\leq1,
\nonumber\\
&\bar A_c^{(j)}\zeta
+\bar B_c^{(j)}\bar\xi
=\bar b_c^{(j)}
\Big\}.
\label{eq:full-cover}
\end{align}
Then (i)
$\Theta_{\rm pcz}^{(j)}
=\mathcal M_{\theta,p}
\cap\Theta_{\rm cz}^{(j)}$, and (ii)
$\Theta_r^{(j)}
\subseteq\Theta_{\rm pcz}^{(j)}
\subseteq\Theta_{\rm cz}^{(j)}$,
so
$\theta^\star\in\Theta_{\rm pcz}^{(j)}$
on the event of
Theorem~\ref{thm:reduced-set}(iv).
\end{lemma}

\begin{proof}
We apply Lemma~1 of \cite{11370532}
for intersection of two CMZs, with
$\mathcal M^1=\Theta_{\rm cz}^{(j)}$,
whose coefficients are $(\zeta,\xi)$,
and $\mathcal M^2=\mathcal M_{\theta,p}$
in \eqref{eq:CMZ-prior}, which has
coefficients $\zeta^p$ and no equality
constraint.

The linking equality becomes
$\sum_sG_{\theta,s}^{(j)}\zeta_s
-\sum_iG_{\theta,p,i}\zeta_i^p
=C_{\theta,p}-C_\theta^{(j)}$.
Since $\{E_s\}$ is an orthonormal basis,
this matrix equality is equivalent to
its $q_\theta$ inner products with
the $E_s$, and by
\eqref{eq:cover-generators}
the $s$-th one is
$w_s\zeta_s-\sum_i[P]_{si}\zeta_i^p
=[c_p]_s-c_s$.
Stacking these below
$A_c^{(j)}\zeta+B_c^{(j)}\xi=b_c^{(j)}$
gives \eqref{eq:full-cover}, which
proves (i).

For (ii),
$\Theta_r^{(j)}
=\mathcal M_{\theta,p}\cap
\Theta_{\rm s}^{(j)}
\subseteq
\mathcal M_{\theta,p}\cap
\Theta_{\rm cz}^{(j)}$
by \eqref{eq:stat-model-set} and
{Lemma~\ref{lem:cover}(iii)}.
The second inclusion is immediate
from (i), and
$\theta^\star\in\Theta_r^{(j)}$
on the stated event.
\end{proof}

The surviving models now refine the covariance certificate: we seek one matrix envelope of $\widehat\Sigma(\theta)$, given by Lemma~\ref{lem:residual-identity}(ii), over the cover $\Theta_{\rm pcz}^{(j)}\supseteq\Theta_r^{(j)}$.

\begin{theorem}
\label{thm:future-cov-refinement}\label{thm:reduced-cov}
Let $\Sigma_{\rm true}\preceq\Sigma^{(j)}$, and write \eqref{eq:full-cover} as
\begin{align}
\Theta_{\rm pcz}^{(j)}
=
\Big\{
C_\theta^{(j)}+\sum_{s=1}^{q_\theta}G_{\theta,s}^{(j)}\chi_s:\
\|\chi\|_\infty\leq1,\ \mathcal A\chi=\bar b_c^{(j)}
\Big\},
\label{eq:cover-stacked}
\end{align}
with $\chi:=\col(\zeta,\bar\xi)\in\mathbb R^{n_\chi}$ and
$\mathcal A:=[\bar A_c^{(j)}\ \ \bar B_c^{(j)}]
\in\mathbb R^{n_e\times n_\chi}$. Let
\begin{align*}
F_0(\Gamma)
&:=
\begin{bmatrix}
\Gamma-\widehat\Sigma_{\rm LS}
&
(\theta_{\rm LS}-C_\theta^{(j)})Q_T^{1/2}\\
\star & I_{n+m}
\end{bmatrix},
\\
F_\ell
&:=
\begin{bmatrix}
0 & -G_{\theta,\ell}^{(j)}Q_T^{1/2}\\
\star & 0
\end{bmatrix},
\quad \ell\leq q_\theta,
\end{align*}
$F_\ell:=0$ for $\ell>q_\theta$, $H\succ0$, and consider, in the symmetric variables $\Gamma$, $X_\ell$, $M_k$,
\begin{align}
\min\ \ &\tr(H\Gamma)
\nonumber\\
\mathrm{s.t.}\ \
&F_0(\Gamma)+\sum_{k=1}^{n_e}[\bar b_c^{(j)}]_kM_k
-\sum_{\ell=1}^{n_\chi}X_\ell\succeq0,
\nonumber\\
&X_\ell\succeq\pm\Big(F_\ell-\sum_{k=1}^{n_e}\mathcal A_{k\ell}M_k\Big),
\quad \ell=1,\ldots,n_\chi,
\nonumber\\
&\Gamma\preceq\alpha_T\Sigma^{(j)} .
\label{eq:reduced-cov-SDP}
\end{align}
{Feasibility is sufficient for a contracting envelope.} If \eqref{eq:reduced-cov-SDP} is feasible, let $\Gamma_j^\star$ be any
feasible point and set
$\Sigma^{(j+1)}:=\alpha_T^{-1}\Gamma_j^\star$. Otherwise, set
$\Sigma^{(j+1)}:=\Sigma^{(j)}$. Then, on $\mathcal E_{\rm all}$,
\begin{align}
\Sigma_{\rm true}
\preceq
\Sigma^{(j+1)}
\preceq
\Sigma^{(j)}.
\label{eq:cov-refinement-nesting}
\end{align}
\end{theorem}

\begin{proof}
If \eqref{eq:reduced-cov-SDP} is infeasible, the claim is the standing
assumption. Otherwise, let $(\Gamma,\{X_\ell\},\{M_k\})$ be feasible and
let $\theta\in\Theta_{\rm pcz}^{(j)}$ be generated by an admissible
$\chi$ in \eqref{eq:cover-stacked}.

 Let
$\theta(\chi):=C_\theta^{(j)}+\sum_{s=1}^{q_\theta}G_{\theta,s}^{(j)}\chi_s$.
By Lemma~\ref{lem:residual-identity}(ii),
$\tfrac1T\widetilde W(\theta)\widetilde W(\theta)^\top
=\widehat\Sigma_{\rm LS}
+(\theta_{\rm LS}-\theta)Q_T(\theta_{\rm LS}-\theta)^\top$, so by the
Schur complement,
$\tfrac1T\widetilde W(\theta(\chi))\widetilde W(\theta(\chi))^\top
\preceq\Gamma$ is 
\begin{align}
F_0(\Gamma)+\sum_{\ell=1}^{n_\chi}\chi_\ell F_\ell\succeq0 .
\label{eq:envelope-affine}
\end{align}
{It must hold for every admissible $\chi$, i.e., with $\|\chi\|_\infty\leq1$ and $\mathcal A\chi=\bar b_c^{(j)}$. Dropping the equalities would ignore the cuts and the prior.}

{To this end, for an admissible}
$\chi$, each scalar
$[\bar b_c^{(j)}]_k-\sum_\ell\mathcal A_{k\ell}\chi_\ell$ is zero, so
$\sum_kM_k\big([\bar b_c^{(j)}]_k-\sum_\ell\mathcal A_{k\ell}\chi_\ell\big)=0$
for arbitrary symmetric $M_k$. Adding this zero matrix to the left-hand
side of \eqref{eq:envelope-affine} and collecting terms gives
\begin{align}
F_0(\Gamma)+\sum_{\ell=1}^{n_\chi}\chi_\ell F_\ell
=
\widetilde F_0+\sum_{\ell=1}^{n_\chi}\chi_\ell\widetilde F_\ell ,
\label{eq:s-procedure-rearrangement}
\end{align}
with
$\widetilde F_0:=F_0(\Gamma)+\sum_k[\bar b_c^{(j)}]_kM_k$ and
$\widetilde F_\ell:=F_\ell-\sum_k\mathcal A_{k\ell}M_k$. {This identity holds only for admissible $\chi$, and $\widetilde F_0$, $\widetilde F_\ell$ are the matrices in the first two constraints of \eqref{eq:reduced-cov-SDP}.}

By the second
constraint of \eqref{eq:reduced-cov-SDP},
$X_\ell\succeq\pm\widetilde F_\ell$, so for $|\chi_\ell|\leq1$,
\(
X_\ell+\chi_\ell\widetilde F_\ell
=
\tfrac{1+\chi_\ell}{2}(X_\ell+\widetilde F_\ell)
+
\tfrac{1-\chi_\ell}{2}(X_\ell-\widetilde F_\ell)
\succeq0 ,
\)
that is, $\chi_\ell\widetilde F_\ell\succeq-X_\ell$. Hence, the
right-hand side of \eqref{eq:s-procedure-rearrangement} is bounded below
by $\widetilde F_0-\sum_\ell X_\ell$, which is positive semidefinite by
the first constraint of \eqref{eq:reduced-cov-SDP}. Thus,
\eqref{eq:envelope-affine} holds for every admissible $\chi$, and
$\tfrac1T\widetilde W(\theta)\widetilde W(\theta)^\top\preceq\Gamma$ for
every $\theta\in\Theta_{\rm pcz}^{(j)}$. With $M_k=0$ the same argument
certifies \eqref{eq:envelope-affine} over the whole box, i.e., over the
matrix zonotope without constraints. The program optimizes over $M_k$
and can therefore only improve on that bound.
On the stated event,
$\theta^\star\in\Theta_r^{(j)}\subseteq\Theta_{\rm pcz}^{(j)}$ by
Lemma~\ref{lem:full-cover}(ii), so
$\tfrac1T\widetilde W^\star\widetilde W^{\star\top}\preceq\Gamma$, and
Lemma~\ref{lem:cov-lower} gives
$\alpha_T\Sigma_{\rm true}
\preceq\tfrac1T\widetilde W^\star\widetilde W^{\star\top}$. Hence,
$\Sigma_{\rm true}\preceq\alpha_T^{-1}\Gamma=\Sigma^{(j+1)}$, and the
last constraint of \eqref{eq:reduced-cov-SDP} gives
$\Sigma^{(j+1)}\preceq\Sigma^{(j)}$.
\end{proof}

\begin{figure}[t]
\centering
\includegraphics{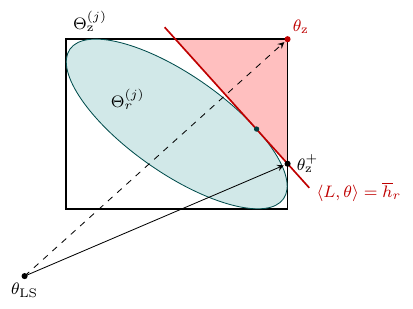}
\caption{One cut-generation step of
Remark~\ref{rem:cut-selection} in the coordinates
$\theta^\top x$ for $n+m=2$ and $Q_T=I$.
{The worst vertex $\theta_{\rm z}$ of $\Theta_{\rm z}^{(j)}$ sets the support radius (dashed). The cut $\langle L,\theta\rangle\leq\overline h_r$, directed from $\theta_{\rm LS}$ toward $\theta_{\rm z}$, touches $\Theta_{\rm s}^{(j)}$ and removes the shaded corner; the new worst vertex $\theta_{\rm z}^{+}$ gives a smaller radius (solid).}}
\label{fig:rank-one-cut}
\end{figure}

\begin{remark}
\label{rem:cut-selection}
By Lemma~\ref{lem:residual-identity}(ii), $x^\top\widehat\Sigma(\theta)x=x^\top\widehat\Sigma_{\rm LS}x+\|a(\theta;x)\|_2^2$ with $a(\theta;x):=Q_T^{1/2}(\theta_{\rm LS}-\theta)^\top x$, so the envelope of Theorem~\ref{thm:future-cov-refinement} in direction $x$ depends on the cover only through
\(
h_\Theta(x)
:=\max_{\theta\in\Theta}\|a(\theta;x)\|_2
=\max_{\|u\|_2=1}\max_{\theta\in\Theta}
\langle x\,u^\top Q_T^{1/2},\theta_{\rm LS}-\theta\rangle,
\)
i.e., through its support in rank-one directions. For a dominant eigenvector $x$ of $H$, we take the worst vertex $\theta_{\rm z}$ of the current cover, found without enumeration by alternating a linear program in $\theta$ with $u\leftarrow a(\theta;x)/\|a(\theta;x)\|_2$, and cut along $L=-x\,a_{\rm z}^\top Q_T^{1/2}$, $a_{\rm z}:=a(\theta_{\rm z};x)$. By Cauchy--Schwarz, the cut removes $\theta_{\rm z}$ whenever $\|a_{\rm z}\|_2>h_{\Theta_{\rm s}^{(j)}}(x)$ (Fig.~\ref{fig:rank-one-cut}), and the step is repeated up to a budget $n_c$. {The closed-form supports of Lemma~\ref{lem:cover}(iv) give such cuts with no SDP. Along $L=\pm E_s$, they intersect the interval hull of $\Theta_{\rm sn}^{(j)}$ with those of the other instruments.}
\end{remark}

\begin{remark}
\label{rem:init-vs-refined-cov}
Program \eqref{eq:reduced-cov-SDP} has $1+2n_\chi$ LMIs of size $2n+m$, $n_\chi=q_\theta+n_c+q_{\theta,p}$, and one $n\times n$ LMI. Its size does not depend on $T$.
\end{remark}

\begin{remark}
\label{rem:sn-envelope}
A complementary candidate bounds $\Theta_{\rm sn}^{(j)}\supseteq\Theta_r^{(j)}$ in closed form. By Lemma~\ref{lem:selfnormalized-MZ}, $(\theta_{\rm LS}-\theta)Q_T^{1/2}={\Omega_q}+C_j\Delta M_q$ on $\Theta_{\rm sn}^{(j)}$, with $\|\Delta\|_2\le1$, ${\Omega_q}:=(\theta_{\rm LS}-\theta_{\rm IV})Q_T^{1/2}$ and
$M_q:=D_\psi\bar\Phi^{-1}Q_T^{1/2}$. For this single full-block uncertainty Petersen's lemma is lossless and gives, for every $\tau>\lammax(M_q^\top M_q)$, the envelope
\begin{align}
& \Gamma_\tau=\tau\Sigma^{(j)}+R_\tau,
\nonumber \\ &
R_\tau:=\widehat\Sigma_{\rm LS}
+{\Omega_q}\big(I_{n+m}-\tau^{-1}M_q^\top M_q\big)^{-1}{\Omega_q^\top}.
\label{eq:Gamma-tau}
\end{align}
Any $\Gamma_\tau\preceq\alpha_T\Sigma^{(j)}$ may replace $\Gamma_j^\star$ in Theorem~\ref{thm:future-cov-refinement}, and the candidate with smaller $\tr(H\Gamma)$ is used. Neither dominates the other. Admissibility requires $\lammax(M_q^\top M_q)<\tau<\alpha_T$, and then the recursion $\Sigma^{(j+1)}=\alpha_T^{-1}\Gamma_\tau$ decreases to $\Sigma_\tau^\infty:=(\alpha_T-\tau)^{-1}R_\tau$ at the geometric rate $\tau/\alpha_T$. Since $R_\tau$ does not depend on $\Sigma^{(j)}$, this limit needs neither recursion nor prior: the proof of Lemma~\ref{lem:iv} gives
$\bar N=\Sigma_{\rm true}^{1/2}\Delta^\star D_\psi$, so \eqref{eq:Gamma-tau} with $\Sigma_{\rm true}$ yields $\alpha_T\Sigma_{\rm true}\preceq\tau\Sigma_{\rm true}+R_\tau$ on $\mathcal E_{\rm all}$, i.e.,
\begin{align}
\Sigma_{\rm true}\preceq\Sigma_\tau^\infty=\frac{R_\tau}{\alpha_T-\tau},
\quad \lammax(M_q^\top M_q)<\tau<\alpha_T .
\label{eq:one-shot}
\end{align}
When $\lammax(M_q^\top M_q)\geq\alpha_T$, i.e.\ with little data, only the prior and \eqref{eq:reduced-cov-SDP} remain. {Every instrument gives its own valid candidates \eqref{eq:Gamma-tau} and \eqref{eq:one-shot}.} {Several instruments are combined by the S-procedure. With $Z:=(\theta_{\rm LS}-\theta)Q_T^{1/2}$, $P^{(i)}:=(M_q^{(i)\top}M_q^{(i)})^{-1}$ and $\Omega^{(i)}:=(\theta_{\rm LS}-\theta_{\rm IV}^{(i)})Q_T^{1/2}$, the set $\Theta_{\rm sn}^{(i,j)}$ is the quadratic matrix inequality $(Z-\Omega^{(i)})P^{(i)}(Z-\Omega^{(i)})^\top\preceq\Sigma^{(j)}$. If $\Gamma$ and $a_i\ge0$ satisfy
\begin{align}
& \begin{bmatrix}
\Gamma-\widehat\Sigma_{\rm LS}&0\\0&-I_{n+m}
\end{bmatrix} \nonumber \\ &
-\sum_{i=1}^{N_\psi}a_i
\begin{bmatrix}
\Sigma^{(j)}-\Omega^{(i)}P^{(i)}\Omega^{(i)\top}&\Omega^{(i)}P^{(i)}\\
\star&-P^{(i)}
\end{bmatrix}
\succeq0,
\label{eq:sproc}
\end{align}
then multiplying by $[I_n\;\,Z]$ on the left and its transpose on the right gives $\widehat\Sigma(\theta)\preceq\Gamma$ on $\bigcap_i\Theta_{\rm sn}^{(i,j)}\supseteq\Theta_r^{(j)}$. This is one LMI of size $2n+m$. For $N_\psi=1$ and $a_1=\tau$, a Schur complement recovers \eqref{eq:Gamma-tau}, and for $N_\psi\ge2$ it exploits that the instrument sets intersect in a smaller set.}
\end{remark}

\subsection{{Iterative Refinement and Finite- and Large-Data Properties}}

Each certificate determines a refined model set through \eqref{eq:reduced-set}, and each refined set produces a candidate certificate through \eqref{eq:reduced-cov-SDP} or \eqref{eq:Gamma-tau}. Algorithm~\ref{alg:iterative-refinement} combines them.

\begin{algorithm}[t]
\caption{Iterative Model-Set and Covariance Refinement}
\label{alg:iterative-refinement}
\textbf{Initialize:} \(\Sigma^{(0)}\) from
Lemma~\ref{lem:init-cert-prior}, \(H\succ0\), cut budget \(n_c\).
\textbf{For} \(j=0,1,\ldots\):

1) \emph{Statistical set.} With \(C_j=(\Sigma^{(j)})^{1/2}\), form
\(\Theta_{\rm s}^{(j)}\) in \eqref{eq:stat-model-set} and
\(\Theta_r^{(j)}=\mathcal M_{\theta,p}\cap\Theta_{\rm s}^{(j)}\).

2) \emph{Cover.} Build \(\Theta_{\rm cz}^{(j)}\) by Lemma~\ref{lem:cover}, with the cuts of Remark~\ref{rem:cut-selection}, and \(\Theta_{\rm pcz}^{(j)}\) by Lemma~\ref{lem:full-cover}.

3) \emph{Envelope.} Solve \eqref{eq:reduced-cov-SDP}, and compute
\eqref{eq:Gamma-tau} {for each instrument} {and, if $N_\psi\ge2$, the smallest-trace $\Gamma$ satisfying \eqref{eq:sproc}}.

4) \emph{Update.} Among the candidates with
\(\Gamma\preceq\alpha_T\Sigma^{(j)}-\varepsilon_{\rm num}I_n\), where $\varepsilon_{\rm num}$ exceeds the solver tolerance and a post-solve check confirms \(\lammax(\Gamma-\alpha_T\Sigma^{(j)})\le0\), let \(\Gamma_j^\star\) minimize
\(\tr(H\Gamma)\) and set
\(\Sigma^{(j+1)}:=\alpha_T^{-1}\Gamma_j^\star\). If there is none, set
\(\Sigma^{(j+1)}:=\Sigma^{(j)}\).

{5) \emph{Consistency.} If} {$\beta_T\Sigma^{(j)}-\widehat\Sigma_{\rm LS}\not\succeq0$} {, the cover is empty, or
$\Theta_r^{(j)}=\emptyset$, stop: the data are inconsistent with \eqref{eq:syst}, \eqref{eq:CMZ-prior}.}

{6)} Stop if the trace decrease is below a tolerance. Otherwise
\(j\leftarrow j+1\).
\end{algorithm}

{Since each accepted certificate is smaller than the previous one, the constraints only tighten as the iteration proceeds.}

\begin{lemma}
\label{lem:finite-T-convergence}
Fix $T$, and let
Algorithm~\ref{alg:iterative-refinement} start from a valid
\(\Sigma^{(0)}\) with
\(\Theta_r^{(0)}\neq\emptyset\). On $\mathcal E_{\rm all}$,
\begin{align}
\Theta_r^{(j+1)}\subseteq\Theta_r^{(j)},
\,\,\,
\Sigma_{\rm true}\preceq\Sigma^{(j+1)}\preceq\Sigma^{(j)},
\,\,\,
\theta^\star\in\Theta_r^{(j)},
\quad
\forall j\ge0,
\label{eq:finite-T-nesting}
\end{align}
\(\Sigma^{(j)}\) converges to some
\(\Sigma_T^{(\infty)}\) with
\(\Sigma_{\rm true}\preceq\Sigma_T^{(\infty)}\preceq\Sigma^{(0)}\), and
\(\Theta_{r,T}^{(\infty)}:=\bigcap_{j\ge0}\Theta_r^{(j)}\) is nonempty,
convex, and compact, with \(\theta^\star\in\Theta_{r,T}^{(\infty)}\).
\end{lemma}

\begin{proof}
At iteration \(j\), Step~1 imposes the two constraints in
\eqref{eq:reduced-set}. If \(\Sigma^{(j+1)}\preceq\Sigma^{(j)}\), then
\((\Sigma^{(j+1)})^{-1}\succeq(\Sigma^{(j)})^{-1}\), so the new
self-normalized constraint implies the old one, {for every instrument}. The prior is unchanged.
Hence no excluded model can re-enter and
\(
\Theta_r^{(j+1)}\subseteq\Theta_r^{(j)}.
\)

By Theorem~\ref{thm:future-cov-refinement} and
Remark~\ref{rem:sn-envelope}, every accepted covariance candidate
satisfies
\(
\Sigma_{\rm true}
\preceq
\Sigma^{(j+1)}
\preceq
\Sigma^{(j)},
\)
while, if no candidate is accepted,
\(\Sigma^{(j+1)}=\Sigma^{(j)}\). Thus, the second nesting relation in
\eqref{eq:finite-T-nesting} holds at every iteration.

Moreover, on the joint confidence event,
Theorem~\ref{thm:reduced-set}(iv) guarantees that the true model
\(\theta^\star\) satisfies both statistical constraints whenever
\(\Sigma_{\rm true}\preceq\Sigma^{(j)}\). Since
\(\theta^\star\in\mathcal M_{\theta,p}\), it is never removed by the
refinement, and therefore
\(
\theta^\star\in\Theta_r^{(j)},
\, \forall j\geq0.
\)

For the limits, \(\{\Sigma^{(j)}\}\) is nonincreasing and bounded below by \(\Sigma_{\rm true}\). Hence, for every
\(x\in\mathbb R^n\),
\(
x^\top\Sigma^{(j)}x
\)
is nonincreasing and bounded below by
\(x^\top\Sigma_{\rm true}x\), and therefore converges. Since the
dimension is finite, there exists
\(\Sigma_T^{(\infty)}\succeq0\) such that
\(
\Sigma^{(j)}\to\Sigma_T^{(\infty)}.
\)
Passing to the limit gives
\(
\Sigma_{\rm true}
\preceq
\Sigma_T^{(\infty)}
\preceq
\Sigma^{(0)}.
\)
Finally, the sets \(\Theta_r^{(j)}\) are nested, convex, and compact,
and each contains \(\theta^\star\). Hence, their intersection
\(
\Theta_{r,T}^{(\infty)}
:=
\bigcap_{j\geq0}\Theta_r^{(j)}
\)
is nonempty, convex, and compact, with
\(\theta^\star\in\Theta_{r,T}^{(\infty)}\).
\end{proof}

Iteration alone cannot remove finite-sample uncertainty. We next let $T\to\infty$.

\begin{assumption}
\label{ass:asymptotic-informativity}
As $T\to\infty$, in probability,
\(
\bar\Phi_T \to \Phi_\infty
\) with $\sigma_{\min}(\Phi_\infty)>0$,
\(
Q_T \to Q_\infty\succ0
\), and
\(
T^{-1}V_{\psi,T} \to R_\psi
\) with $0\prec R_\psi\prec\infty$.
For large-data statements, $\delta_c(T),\delta_s(T),\delta_1(T)\to0$ with
$\log(1/\delta_i(T))=O(\log T)$.
\end{assumption}

\begin{lemma}
\label{lem:large-T-consistency}
Under Assumptions~\ref{noises}--\ref{ass:asymptotic-informativity}, for every fixed $j$,
\begin{align}
\sup_{\theta\in\Theta_r^{(j)}}
\|\theta-\theta^\star\|_2
&=
O_{\mathbb P}(\sqrt{\log T/T}),
\label{eq:model-consistency-rate}\\
\sup_{\theta\in\Theta_r^{(j)}}
\|N(\theta)-\bar N\|_2
&=
O_{\mathbb P}(\sqrt{\log T/T}),
\label{eq:N-consistency-rate}
\end{align}
so \(\Theta_r^{(j)}\to\{\theta^\star\}\) in probability.
\end{lemma}

\begin{proof}
Since \(\Theta_r^{(j)}\subseteq\Theta_{\rm sn}^{(j)}\), it suffices to
bound the radius of the self-normalized set.  By
\eqref{eq:theta-SN-MZ}, every \(\theta\in\Theta_{\rm sn}^{(j)}\) satisfies
\(\theta_{\rm IV}-\theta=C_j\Delta D_\psi\bar\Phi^{-1}\) for some
\(\|\Delta\|_2\le1\), so
\begin{align}
\|\theta-\theta_{\rm IV}\|_2
\le
\|(\Sigma^{(j)})^{1/2}\|_2\,\|D_\psi\|_2\,\|\bar\Phi^{-1}\|_2 .
\label{eq:sn-radius}
\end{align}
On the confidence event, $\theta^\star\in\Theta_{\rm sn}^{(j)}$ and
therefore satisfies the same radius bound \eqref{eq:sn-radius} as any
$\theta\in\Theta_{\rm sn}^{(j)}$. Hence, by the triangle inequality,
\begin{align} \label{new1}
\|\theta-\theta^\star\|_2
\leq
2
\|(\Sigma^{(j)})^{1/2}\|_2
\|D_\psi\|_2
\|\bar\Phi^{-1}\|_2.
\end{align}
By
Theorem~\ref{thm:future-cov-refinement},
\(\Sigma_{\rm true}\preceq\Sigma^{(j)}\preceq\Sigma^{(0)}\). For a fixed
prior, Assumption~\ref{ass:asymptotic-informativity} and
Lemma~\ref{lem:init-cert-prior} give
\(\|\Sigma^{(0)}\|_2=O_{\mathbb P}(1)\). Hence,
\(\|(\Sigma^{(j)})^{1/2}\|_2=O_{\mathbb P}(1)\), while
\(\|\bar\Phi^{-1}\|_2=O_{\mathbb P}(1)\).  Under the
standing excitation assumptions of Assumption~5, we have \(\|V_\psi^{1/2}\|_2=O_{\mathbb P}(\sqrt
T)\) and hence \(\gamma(\delta_s)=O_{\mathbb P}(\sqrt{\log T})\). Therefore, using \(D_\psi=\frac{\sqrt2\,\gamma(\delta_s)}{T}V_\psi^{1/2}\), one has 
\(\|D_\psi\|_2=O_{\mathbb P}(\sqrt{\log T/T})\).  Using this in \eqref{new1} gives
\eqref{eq:model-consistency-rate}.  For
\eqref{eq:N-consistency-rate}, \(N(\theta)-\bar N
=-(\theta-\theta^\star)\bar\Phi\) and \(\|\bar\Phi\|_2\) is bounded, so the
same rate applies.  Finally, the confidence-event probability tends to one by
Assumption~\ref{ass:asymptotic-informativity}. On that event
\(\theta^\star\in\Theta_r^{(j)}\) and the maximum distance to it vanishes,
which proves the stated convergence in probability.
\end{proof}

The covariance counterpart of Lemma~\ref{lem:large-T-consistency} needs an initialization margin.

\begin{corollary}
\label{cor:cov-recovery}
Let Assumption~\ref{ass:asymptotic-informativity} hold, with fixed $n_c$ and $H\succ0$ and \eqref{eq:reduced-cov-SDP} solved to optimality, and suppose that, for some $c_0>0$ and with probability tending to one,
\begin{align}
\alpha_T\Sigma^{(0)}-\tfrac1T\widetilde W^\star\widetilde W^{\star\top}\succeq c_0I_n .
\label{eq:init-margin}
\end{align}
Then, for every fixed $j$,
$\Prob[\Sigma_{\rm true}\preceq\Sigma^{(j+1)}]\to1$ and
$\|\Sigma^{(j+1)}-\Sigma_{\rm true}\|_2=O_{\mathbb P}(\sqrt{\log T/T})$, and the same holds for $\Sigma_T^{(\infty)}$.
\end{corollary}

\begin{proof}
Let
\(
r_T:=\sqrt{\frac{\log T}{T}}.
\)
{By Lemma~\ref{lem:finite-T-convergence},
$\Sigma_{\rm true}\preceq\Sigma^{(j+1)}\preceq\Sigma^{(1)}$ for every $j\ge0$,
so $\|\Sigma^{(j+1)}-\Sigma_{\rm true}\|_2\le\|\Sigma^{(1)}-\Sigma_{\rm true}\|_2$.
It therefore suffices to prove the rate for $\Sigma^{(1)}$, and we take $j=0$
in the remainder of the proof.}
The radius argument in Lemma~\ref{lem:large-T-consistency} applies already
to the self-normalized set
\(\Theta_{\rm sn}^{(j)}\), and
\(
\Theta_{\rm s}^{(j)}
\subseteq
\Theta_{\rm sn}^{(j)}.
\)
Hence,
\(
\sup_{\theta\in\Theta_{\rm s}^{(j)}}
\|\theta-\theta^\star\|_2
=
O_{\mathbb P}(r_T).
\)
Therefore, every coordinate width in
\eqref{eq:cover-generators} is \(O_{\mathbb P}(r_T)\), and so
\(
\|G_{\theta,s}^{(j)}\|_2
=
O_{\mathbb P}(r_T).
\)
Thus, both the interval hull and its CMZ tightening have diameter
\(O_{\mathbb P}(r_T)\) around \(\theta^\star\). The fixed number of supporting cuts can only reduce this cover.

By Lemma~\ref{lem:residual-identity}(ii),
\(
\widehat\Sigma(\theta)
:=
\frac1T\widetilde W(\theta)\widetilde W(\theta)^\top
=
\widehat\Sigma_{\rm LS}
+
(\theta_{\rm LS}-\theta)Q_T(\theta_{\rm LS}-\theta)^\top .
\)
Since \(\|Q_T\|_2=O_{\mathbb P}(1)\){,
$\|\theta_{\rm LS}-\theta^\star\|_2=O_{\mathbb P}(1)$,} and the cover shrinks to
\(\theta^\star\), {one has}
\(
\sup_{\theta\in\Theta_{\rm cz}^{(j)}}
\|\widehat\Sigma(\theta)-\widehat\Sigma(\theta^\star)\|_2
=
O_{\mathbb P}(r_T).
\)
Moreover, every uncertain block \(F_\ell\) in
\eqref{eq:reduced-cov-SDP} is linear in
\(G_{\theta,\ell}^{(j)}\), and hence is
\(O_{\mathbb P}(r_T)\).
{Let $\epsilon_T:=\sum_{\ell\le q_\theta}\|F_\ell\|_2=O_{\mathbb P}(r_T)$
and {recall $\Omega_q=(\theta_{\rm LS}-C_\theta^{(0)})Q_T^{1/2}$ from Remark~\ref{rem:sn-envelope}, since $C_\theta^{(0)}=\theta_{\rm IV}$,} so that $\|{\Omega_q}\|_2=O_{\mathbb P}(1)$.
With $M_k=0$ and $X_\ell=\|F_\ell\|_2I$, the second constraint of
\eqref{eq:reduced-cov-SDP} holds, and by the Schur complement the first
constraint holds (for $\epsilon_T<1$) at
\(
\widetilde\Gamma_0
:=\widehat\Sigma_{\rm LS}+\epsilon_TI_n+(1-\epsilon_T)^{-1}{\Omega_q\Omega_q^\top}
=\widehat\Sigma\big(C_\theta^{(0)}\big)+\epsilon_TI_n+\tfrac{\epsilon_T}{1-\epsilon_T}{\Omega_q\Omega_q^\top} .
\)
{A constrained-zonotope center need not satisfy its coefficient
equalities, so no membership of $C_\theta^{(0)}$ in
$\Theta_{\rm pcz}^{(0)}$ is used. Instead,
$C_\theta^{(0)}=\theta_{\rm IV}$, and, on $\mathcal E_{\rm all}$,
$\theta^\star\in\Theta_{\rm sn}^{(0)}$. Hence,
\eqref{eq:sn-radius} gives
$\|C_\theta^{(0)}-\theta^\star\|_2=O_{\mathbb P}(r_T)$.
The residual identity, $\|Q_T\|_2=O_{\mathbb P}(1)$, and
$\|\theta_{\rm LS}-\theta^\star\|_2=O_{\mathbb P}(1)$ then imply
$\|\widehat\Sigma(C_\theta^{(0)})-\widehat\Sigma(\theta^\star)\|_2
=O_{\mathbb P}(r_T)$, and therefore
$\|\widetilde\Gamma_0-\widehat\Sigma(\theta^\star)\|_2
=O_{\mathbb P}(r_T)$.} By
\eqref{eq:init-margin} and $r_T\to0$, with probability tending to one
$\widetilde\Gamma_0\preceq\alpha_T\Sigma^{(0)}$, so $\widetilde\Gamma_0$ is
feasible for \eqref{eq:reduced-cov-SDP}. Hence a candidate exists, and the
update $\Gamma_0^\star$ accepted in Step~4 satisfies
$\operatorname{tr}(H\Gamma_0^\star)\le\operatorname{tr}(H\widetilde\Gamma_0)$.
On $\mathcal E_{\rm all}$, $\theta^\star$ belongs to both $\Theta_{\rm pcz}^{(0)}$
and $\Theta_{\rm sn}^{(0)}$, so every candidate dominates
$\widehat\Sigma(\theta^\star)$, i.e.\
$\Gamma_0^\star-\widehat\Sigma(\theta^\star)\succeq0$. Therefore,
\(
\lambda_{\min}(H)\,\big\|\Gamma_0^\star-\widehat\Sigma(\theta^\star)\big\|_2
\le\operatorname{tr}\!\big(H(\Gamma_0^\star-\widehat\Sigma(\theta^\star))\big)
\le\operatorname{tr}(H)\,\big\|\widetilde\Gamma_0-\widehat\Sigma(\theta^\star)\big\|_2 .
\)}
Consequently,
\(
\Gamma_0^\star
=
\frac1T
\widetilde W^\star\widetilde W^{\star\top}
+
O_{\mathbb P}(r_T)
\)
in spectral norm.

It remains to relate the empirical covariance to
\(\Sigma_{\rm true}\). From the pivotal construction of
Lemma~\ref{lem:cov-lower},
\(
\alpha_T\Sigma_{\rm true}
\preceq
\frac1T\widetilde W^\star\widetilde W^{\star\top}
\preceq
\beta_T\Sigma_{\rm true}.
\)
For the Gaussian matrices defining the pivotal quantiles in
\eqref{eq:pivotal-quantiles}, standard extreme-singular-value concentration
and the confidence allocation of Assumption~\ref{ass:asymptotic-informativity}
give
\(
1-\alpha_T=O(\sqrt{\log T/T}),
\,
\beta_T-1=O(\sqrt{\log T/T}).
\)
Hence,
\(
\big\|
\frac1T\widetilde W^\star\widetilde W^{\star\top}
-\Sigma_{\rm true}
\big\|_2
=
O_{\mathbb P}(r_T).
\)
Combining the last two displays gives
\(
\|\Gamma_0^\star-\Sigma_{\rm true}\|_2
=
O_{\mathbb P}(r_T).
\)
Since also
\(
\alpha_T^{-1}=1+O(r_T),
\)
the accepted first update
\(
\Sigma^{(1)}
=
\alpha_T^{-1}\Gamma_0^\star
\)
satisfies
\(
\|\Sigma^{(1)}-\Sigma_{\rm true}\|_2
=
O_{\mathbb P}(r_T).
\)
The containment
\(\Sigma_{\rm true}\preceq\Sigma^{(j+1)}\) follows from
Theorem~\ref{thm:future-cov-refinement}.
This proves the rate for $\Sigma^{(1)}$, and hence for every
$\Sigma^{(j+1)}$ by the first step, whether or not later candidates are accepted.

Finally, Lemma~\ref{lem:finite-T-convergence} gives
\(
\Sigma_{\rm true}
\preceq
\Sigma_T^{(\infty)}
\preceq
\Sigma^{(j+1)}
\)
for every \(j\). Therefore,
\(
\|\Sigma_T^{(\infty)}-\Sigma_{\rm true}\|_2
\le
\|\Sigma^{(j+1)}-\Sigma_{\rm true}\|_2
=
O_{\mathbb P}(r_T),
\)
which proves the claim.
\end{proof}

The margin \eqref{eq:init-margin} is ensured, for instance, by a positive lower bound on $\mu_p^2 - T^{-1}\|E_p\|_2^2$, since the Young slack in the proof of Lemma~3 is already nonnegative. The coefficient slack alone is insufficient.

\subsection{Quantified Refinement Gain}
\label{sec:refinement-gain}

Theorem~\ref{thm:future-cov-refinement} and Lemma~\ref{lem:finite-T-convergence} establish only nonexpansion, and Lemma~\ref{lem:large-T-consistency} credits refinement with no asymptotic improvement. This subsection quantifies the gain and where it saturates.

\begin{proposition}
\label{prop:envelope-closed-form}
For a nonempty compact
\(\Theta\subseteq\mathbb R^{n\times(n+m)}\), let
\begin{align}
h_\Theta(x)
:=
\max_{\theta\in\Theta}
\big\|
Q_T^{1/2}(\theta_{\rm LS}-\theta)^\top x
\big\|_2 ,
\label{eq:support-radius}
\end{align}
\begin{align}
J_\Theta(x)
:=
\inf_{\Gamma}
\Big\{
x^\top\Gamma x:\;
\widehat\Sigma(\theta)
\preceq\Gamma\
\forall\,\theta\in\Theta
\Big\}.
\label{eq:directional-envelope}
\end{align}
Then, for every \(x\), the nesting
\(
\Theta_r^{(j+1)}\subseteq\Theta_r^{(j)}
\)
implies
\begin{align}
J_{\Theta_r^{(j)}}(x)
-
J_{\Theta_r^{(j+1)}}(x)
=
h_{\Theta_r^{(j)}}(x)^2
-
h_{\Theta_r^{(j+1)}}(x)^2
\geq0 .
\label{eq:directional-refinement-gain}
\end{align}
For $H\succeq0$ with unit eigenvectors $x_i$ and eigenvalues $\lambda_i$, the value $v_H(\Theta)$ of $\min\tr(H\Gamma)$ over the envelopes in \eqref{eq:directional-envelope} satisfies
\begin{align}
\sum_{i}\lambda_ih_\Theta(x_i)^2
\!\le\!
v_H(\Theta)-\tr(H\widehat\Sigma_{\rm LS})
\!\le\!
\tr(H)\!\max_{\|x\|_2=1}\!h_\Theta(x)^2 .
\label{eq:trace-sandwich}
\end{align}
\end{proposition}

\begin{proof}
By Lemma~\ref{lem:residual-identity}(ii), for every
\(\theta\in\Theta\),
\(
\frac1T
\widetilde W(\theta)\widetilde W(\theta)^\top
=
\widehat\Sigma_{\rm LS}
+
(\theta_{\rm LS}-\theta)
Q_T
(\theta_{\rm LS}-\theta)^\top .
\)
Hence, for any \(x\in\mathbb R^n\),
\begin{align}
x^\top
\left(
\frac1T
\widetilde W(\theta)\widetilde W(\theta)^\top
\right)x
=
x^\top\widehat\Sigma_{\rm LS}x
+
\big\|
Q_T^{1/2}
(\theta_{\rm LS}-\theta)^\top x
\big\|_2^2 .
\label{eq:prop-directional}
\end{align}

For any \(\Gamma\) in
\eqref{eq:directional-envelope}, we have
\(
x^\top\Gamma x
\geq
\max_{\theta\in\Theta}
x^\top
\big(
\frac1T
\widetilde W(\theta)\widetilde W(\theta)^\top
\big)x .
\)
Using \eqref{eq:prop-directional} and the definition of
\(h_\Theta(x)\) gives
\(
x^\top\Gamma x
\geq
x^\top\widehat\Sigma_{\rm LS}x
+
h_\Theta(x)^2 .
\)
Taking the infimum over all feasible \(\Gamma\) yields
\begin{align}
J_\Theta(x)
\geq
x^\top\widehat\Sigma_{\rm LS}x
+
h_\Theta(x)^2 .
\label{eq:prop-lower}
\end{align}

We now show that this directional lower bound is achievable
arbitrarily closely. The case \(x=0\) is immediate, so suppose
\(x\neq0\), let
\(
e_1:=x/\|x\|_2
\),
and complete it to an orthogonal matrix
\(
U=[\,e_1\ \ U_\perp\,].
\)
For
\(
S(\theta)
:=
\frac1T
\widetilde W(\theta)\widetilde W(\theta)^\top,
\)
write
\[
U^\top S(\theta)U
=
\begin{bmatrix}
a_\theta & b_\theta^\top\\
b_\theta & C_\theta
\end{bmatrix}.
\]
Since \(\Theta\) is compact and \(S(\theta)\) is continuous,
\(
a^\star:=\max_{\theta\in\Theta}a_\theta,
\,
B:=\max_{\theta\in\Theta}\|b_\theta\|_2,
\,
C:=\max_{\theta\in\Theta}\|C_\theta\|_2
\)
are finite. For any \(\varepsilon>0\), define
\[
\Gamma_\varepsilon
:=
U
\begin{bmatrix}
a^\star+\varepsilon & 0\\
0 & \lambda_\varepsilon I_{n-1}
\end{bmatrix}
U^\top,
\qquad
\lambda_\varepsilon
:=
C+\frac{B^2}{\varepsilon}+1.
\]
For every \(\theta\in\Theta\),
\(a^\star+\varepsilon-a_\theta\geq\varepsilon\), and the Schur
complement gives
\(
\lambda_\varepsilon I_{n-1}
-
C_\theta
-
\frac{b_\theta b_\theta^\top}
{a^\star+\varepsilon-a_\theta}
\succeq I_{n-1}\succ0.
\)
Therefore,
\(
S(\theta)\preceq\Gamma_\varepsilon,
\,
\forall\,\theta\in\Theta,
\)
so \(\Gamma_\varepsilon\) is feasible in
\eqref{eq:directional-envelope}. Moreover,
\(
x^\top\Gamma_\varepsilon x
=
\|x\|_2^2(a^\star+\varepsilon).
\)
By \eqref{eq:prop-directional},
\(
\|x\|_2^2a^\star
=
x^\top\widehat\Sigma_{\rm LS}x
+
h_\Theta(x)^2.
\)
Letting \(\varepsilon\downarrow0\) and combining with
\eqref{eq:prop-lower} gives
\begin{align}
J_\Theta(x)
=
x^\top\widehat\Sigma_{\rm LS}x
+
h_\Theta(x)^2 .
\label{eq:envelope-support}
\end{align}

Now suppose
\(\Theta_1\subseteq\Theta_2\). Since \(h_\Theta(x)\) is defined by
maximization over \(\Theta\), it yields
\(
h_{\Theta_1}(x)
\leq
h_{\Theta_2}(x).
\)
Equation~\eqref{eq:envelope-support} then immediately gives
\(
J_{\Theta_1}(x)
\leq
J_{\Theta_2}(x).
\) Finally, Algorithm~\ref{alg:iterative-refinement} gives
\(
\Theta_r^{(j+1)}
\subseteq
\Theta_r^{(j)}.
\)
Applying \eqref{eq:envelope-support} to the two nested sets and
subtracting cancels the common term
\(x^\top\widehat\Sigma_{\rm LS}x\), yielding \eqref{eq:directional-refinement-gain}.

{For \eqref{eq:trace-sandwich}, every feasible \(\Gamma\) satisfies
\(x_i^\top\Gamma x_i\ge J_\Theta(x_i)\) and
\(\tr(H\Gamma)=\sum_i\lambda_ix_i^\top\Gamma x_i\), which gives the lower bound
by \eqref{eq:envelope-support}. For the upper bound,
\(\Gamma=\widehat\Sigma_{\rm LS}+\max_{\|x\|_2=1}h_\Theta(x)^2I_n\) is feasible by
Lemma~\ref{lem:residual-identity}(ii). The bounds coincide for \(n=1\). For \(n>1\), a single
envelope realizes \eqref{eq:directional-refinement-gain} only up to this gap.}
\end{proof}

{In each direction $x$, the gain from refinement is the drop in the support radius when the self-normalized set is replaced by the refined set. The prior guarantees a gain wherever it is narrower than the self-normalized set.}

\begin{theorem}
\label{thm:strict-gain}
Define the directional refinement gain
\begin{align}
\Delta_j(x)
:=
\frac{
h_{\Theta_{\rm sn}^{(j)}}(x)^2
-
h_{\Theta_r^{(j)}}(x)^2
}{\alpha_T}
\geq0,
\label{eq:gain-exact}
\end{align}
which is positive if and only if no maximizer defining
$h_{\Theta_{\rm sn}^{(j)}}(x)$ lies in $\Theta_r^{(j)}$. Then, for every $x\in\mathbb R^n$:
\begin{enumerate}
\item[{(i)}] with
\(
a(x):=Q_T^{1/2}(\theta_{\rm LS}-\theta_{\rm IV})^\top x
\),
\(
M:=Q_T^{1/2}\bar\Phi^{-\top}D_\psi
\) and
\(
b(x):=C_j^\top x
\),
\begin{align}
& h_{\Theta_{\rm sn}^{(j)}}(x)
=
\max_{\|w\|_2\leq\|b(x)\|_2}
\|a(x)+Mw\|_2
\leq \nonumber \\ &
\|a(x)\|_2+\|M\|_2\|b(x)\|_2.
\label{eq:hsn-closed}
\end{align}
\item[{(ii)}]
if
\(
\|Q_T^{1/2}(\theta_{\rm LS}-\theta)^\top x\|_2
\leq h_p(x)
\) for all $\theta\in\mathcal M_{\theta,p}$, then
\begin{align}
\Delta_j(x)
\geq
\frac{
\big[
h_{\Theta_{\rm sn}^{(j)}}(x)^2-h_p(x)^2
\big]_+
}{\alpha_T},
\label{eq:gain-lower}
\end{align}
so the gain is strict whenever
$h_p(x)<h_{\Theta_{\rm sn}^{(j)}}(x)$.
\end{enumerate}
\end{theorem}

\begin{proof}
For part~{(i)}, every $\theta\in\Theta_{\rm sn}^{(j)}$ satisfies
\(
\theta_{\rm IV}-\theta
=
C_j\Delta D_\psi\bar\Phi^{-1},
\,
\|\Delta\|_2\leq1.
\)
Hence,
\(
Q_T^{1/2}(\theta_{\rm LS}-\theta)^\top x
=
a(x)+M\Delta^\top b(x).
\)
As $\Delta$ ranges over the unit spectral-norm ball,
$\Delta^\top b(x)$ ranges exactly over
\(
\{w:\|w\|_2\leq\|b(x)\|_2\}.
\)
Indeed,
$\|\Delta^\top b(x)\|_2\leq\|b(x)\|_2$, while for any such $w$ and
$b(x)\neq0$,
\(
\Delta^\top
=
\frac{w\,b(x)^\top}{\|b(x)\|_2^2}
\)
has $\|\Delta\|_2\leq1$ and satisfies
$\Delta^\top b(x)=w$. The case $b(x)=0$ is immediate. This proves the
equality in \eqref{eq:hsn-closed}, and the upper bound follows from the
triangle inequality.

For part~{(ii)}, since
\(
\Theta_r^{(j)}
\subseteq
\mathcal M_{\theta,p}\cap\Theta_{\rm sn}^{(j)},
\)
one has
\(
h_{\Theta_r^{(j)}}(x)
\leq
\min\{
h_p(x),\,h_{\Theta_{\rm sn}^{(j)}}(x)
\}.
\)
Substituting this bound into \eqref{eq:gain-exact} gives
\eqref{eq:gain-lower}. The bound isolates the gain guaranteed by the
prior restriction. The exact gain $\Delta_j(x)$ may be larger because
{additional instruments can further reduce the refined set.}
\end{proof}

\begin{remark}
\label{rem:exact-vs-implemented-gain}
The implemented update adds a cover gap
$\epsilon_{\rm cov}^{(j)}(x):=h_{\Theta_{\rm pcz}^{(j)}}(x)^2-h_{\Theta_r^{(j)}}(x)^2\ge0$ and a relaxation gap
$\epsilon_{\rm sdp}^{(j)}(x):=x^\top\Gamma_{\rm alg}^{(j)}x-J_{\Theta_{\rm pcz}^{(j)}}(x)\ge0$, with $x^\top\Gamma_{\rm alg}^{(j)}x-J_{\Theta_r^{(j)}}(x)=\epsilon_{\rm cov}^{(j)}(x)+\epsilon_{\rm sdp}^{(j)}(x)$, so a strict exact gain need not give a strict update. Cuts reduce $\epsilon_{\rm cov}$, and tighter relaxations reduce $\epsilon_{\rm sdp}$.
\end{remark}

An accepted contraction propagates back to the model set as follows.

\begin{theorem}
\label{thm:geometric-contraction}
Let
\(
\kappa_j
:=
\lammax\big(
(\Sigma^{(j)})^{-1/2}\Sigma^{(j+1)}(\Sigma^{(j)})^{-1/2}
\big)\in(0,1],
\)
so that \(\Sigma^{(j+1)}\preceq\kappa_j\Sigma^{(j)}\). Then,
\begin{align}
\Theta_{\rm sn}^{(j+1)}
\subseteq
\theta_{\rm IV}+\sqrt{\kappa_j}\,
\big(\Theta_{\rm sn}^{(j)}-\theta_{\rm IV}\big),
\label{eq:set-contraction}
\end{align}
and, on $\mathcal E_{\rm all}$, for every \(J\geq1\),
\begin{align}
\sup_{\theta\in\Theta_r^{(J)}}\|\theta-\theta^\star\|_2
\;\leq\;
2\Big(\prod_{i=0}^{J-1}\sqrt{\kappa_i}\Big)
\big\|C_0\big\|_2\big\|D_\psi\big\|_2\big\|\bar\Phi^{-1}\big\|_2 .
\label{eq:radius-contraction}
\end{align}
Moreover, on $\mathcal E_{\rm all}$, with $\Theta_{\rm sn}^{\star}$ the
set \eqref{eq:theta-SN-MZ} built with $C_j=\Sigma_{\rm true}^{1/2}$ and
$\rho_J:=\lammax(\Sigma_{\rm true}^{-1/2}\Sigma^{(J)}\Sigma_{\rm true}^{-1/2})\ge1$,
\begin{align}
&\Theta_{\rm sn}^{\star}
\subseteq
\Theta_{\rm sn}^{(J)}
\subseteq
\theta_{\rm IV}+\sqrt{\rho_J}\,
\big(\Theta_{\rm sn}^{\star}-\theta_{\rm IV}\big),
\label{eq:oracle-sandwich}\\
&\prod_{i=0}^{J-1}\kappa_i
\ \ge\
\lammax\!\big((\Sigma^{(0)})^{-1/2}\Sigma_{\rm true}(\Sigma^{(0)})^{-1/2}\big).
\label{eq:kappa-floor}
\end{align}
\end{theorem}

\begin{proof}
Let
\(M(\theta):=(\theta_{\rm IV}-\theta)\bar\Phi D_\psi^{-1}\).
Then,
\(\|C_{j+1}^{-1}M\|_2^2
=\lammax(M^\top(\Sigma^{(j+1)})^{-1}M)\).
From
\(\Sigma^{(j+1)}\preceq\kappa_j\Sigma^{(j)}\)
it follows that
\((\Sigma^{(j+1)})^{-1}
\succeq\kappa_j^{-1}(\Sigma^{(j)})^{-1}\), hence
\(\|C_{j+1}^{-1}M\|_2^2
\geq\kappa_j^{-1}\|C_j^{-1}M\|_2^2\).
Therefore
\(\theta\in\Theta_{\rm sn}^{(j+1)}\), i.e.\
\(\|C_{j+1}^{-1}M(\theta)\|_2\leq1\), implies
\(\|C_j^{-1}M(\theta)\|_2\leq\sqrt{\kappa_j}\).
Since \(\Theta_{\rm sn}^{(j)}\) is the preimage of the unit spectral ball
under the affine map
\(\theta\mapsto C_j^{-1}M(\theta)\) and is therefore centrally symmetric
about \(\theta_{\rm IV}\), the sublevel set at radius
\(\sqrt{\kappa_j}\) is exactly the right-hand side of
\eqref{eq:set-contraction}. Iterating and using
\(\Theta_r^{(J)}\subseteq\Theta_{\rm sn}^{(J)}\),
\(\theta^\star\in\Theta_r^{(J)}\), together with the radius bound
\(\|\theta_{\rm IV}-\theta\|_2
\leq\|C_0\|_2\|D_\psi\|_2\|\bar\Phi^{-1}\|_2\)
valid on \(\Theta_{\rm sn}^{(0)}\), gives
\eqref{eq:radius-contraction}.
{In \eqref{eq:oracle-sandwich}, the first inclusion follows from
$\Sigma_{\rm true}\preceq\Sigma^{(J)}$ and the second from
$\Sigma^{(J)}\preceq\rho_J\Sigma_{\rm true}$ by the argument used for
\eqref{eq:set-contraction}. \eqref{eq:kappa-floor} follows from
$\Sigma_{\rm true}\preceq\Sigma^{(J)}\preceq(\prod_i\kappa_i)\Sigma^{(0)}$.}
\end{proof}

\begin{remark}
\label{rem:refinement-scope}
Refinement preserves the rate $O_{\mathbb P}(\sqrt{\log T/T})$ and improves its constant through $\prod_i\sqrt{\kappa_i}$, which by \eqref{eq:kappa-floor} removes the looseness of the initialization and nothing more. By \eqref{eq:oracle-sandwich}, the refined set is within the factor $\sqrt{\rho_J}$ of the known-covariance set, and $\rho_J\to1$ by Corollary~\ref{cor:cov-recovery}. The prior matters in the opposite, low-data regime ($\lammax(M_q^\top M_q)\ge\alpha_T$), where the gain \eqref{eq:gain-lower} is largest.
\end{remark}

\subsection{Certified Separate Covariance Bounds}
\label{sec:WV-certified}

Part~II~\cite{PartII} needs bounds on \(W\) and \(V\) separately, whereas Theorem~\ref{thm:future-cov-refinement} certifies only \(\Sigma_{\rm true}=W+V+A^\star VA^{\star\top}\). The moving-average structure of \(\widetilde w_k\) separates the two through its lag-one covariance.

\begin{lemma}
\label{lem:lag1-radius}
Let
\(
\widehat\Sigma_1(\theta)
:=
T^{-1}\widetilde W(\theta)S_1\widetilde W(\theta)^\top
\)
and
\(
\Sigma_1
:=
\E[\widetilde w_k\widetilde w_{k+1}^\top]
=
-VA^{\star\top}.
\)
Under Assumptions~\ref{noises}--\ref{ass:asymptotic-informativity},
\begin{align}
\widehat\Sigma_1(\theta^\star)
=
-VA^{\star\top}
+
O_{\mathbb P}(T^{-1/2}),
\label{eq:lag1-radius}
\end{align}
and, for every fixed $j$,
\begin{align}
\sup_{\theta\in\Theta_r^{(j)}}
\big\|
\widehat\Sigma_1(\theta)+VA^{\star\top}
\big\|_2
=
O_{\mathbb P}\big(
\sqrt{{\log T}/{T}}
\big).
\label{eq:lag1-uniform}
\end{align}
\end{lemma}

\begin{proof}
From
\eqref{effective1}, one has
\(
\widetilde w_k
=
\omega_k+\upsilon_{k+1}-A^\star\upsilon_k,
\,
\widetilde w_{k+1}
=
\omega_{k+1}+\upsilon_{k+2}-A^\star\upsilon_{k+1}.
\)
By Assumption~\ref{noises}, all cross terms have zero expectation except
the one involving the common measurement noise \(\upsilon_{k+1}\). Hence,
\[
\Sigma_1
=
\E[\widetilde w_k\widetilde w_{k+1}^\top]
=
\E\!\left[
\upsilon_{k+1}(-A^\star\upsilon_{k+1})^\top
\right]
=
-VA^{\star\top}
\]
is the population lag-one covariance. 
For the finite-sample bound, define
\(
Z_k
:=
\widetilde w_k\widetilde w_{k+1}^\top-\Sigma_1,
\, k=0,\ldots,T-2.
\)
Then, \(\E[Z_k]=0\). Moreover, \(Z_k\) depends only on
\(
(\omega_k,\omega_{k+1},
\upsilon_k,\upsilon_{k+1},\upsilon_{k+2}),
\)
so \(Z_k\) and \(Z_\ell\) are independent whenever
\(|k-\ell|\ge3\). Therefore, splitting the indices into
\(
\mathcal I_q
:=
\{k\in\{0,\ldots,T-2\}:k\equiv q\pmod 3\},
\, q=0,1,2,
\)
produces three subsequences of independent centered matrices.

On the other hand, for fixed unit vectors \(u,v\), we have
\(
u^\top Z_kv
=
(u^\top\widetilde w_k)
(v^\top\widetilde w_{k+1})
-u^\top\Sigma_1v .
\)
The two factors are jointly Gaussian with variances bounded by
\(\|\Sigma_{\rm true}\|_2\). Hence, \(u^\top Z_kv\) is centered
sub-exponential with scale bounded by an absolute constant times
\(\|\Sigma_{\rm true}\|_2\). Bernstein's inequality \cite{tropp2015matrix} on each residue
class, followed by a \(1/4\)-net argument over the two unit spheres and
a union bound over \(q=0,1,2\), therefore gives
\(
\big\|
\frac1T\sum_{k=0}^{T-2}Z_k
\big\|_2
\le
c\,\|\Sigma_{\rm true}\|_2
\sqrt{\frac{n+\log(4/\delta_1)}{T}}
\)
with probability at least \(1-\delta_1\), for \(T\) sufficiently large
relative to \(n+\log(4/\delta_1)\). 
Besides, since
\(
\widehat\Sigma_1(\theta^\star)
=
\frac1T\sum_{k=0}^{T-2}
\widetilde w_k\widetilde w_{k+1}^\top,
\)
we have
\(
\widehat\Sigma_1(\theta^\star)-\Sigma_1
=
\frac1T\sum_{k=0}^{T-2}Z_k-\frac1T\Sigma_1.
\)
The last term is the endpoint effect caused by normalizing the \(T-1\)
lag-one products by \(T\). Since
\(\|\Sigma_1\|_2\le\|\Sigma_{\rm true}\|_2\) and
\(\Sigma_{\rm true}\preceq\Sigma^{(j)}\),
\[
\big\|
\widehat\Sigma_1(\theta^\star)-\Sigma_1
\big\|_2
\le
c\,\|\Sigma^{(j)}\|_2
\sqrt{\frac{n+\log(4/\delta_1)}{T}}
+
O(T^{-1}),
\]
which proves \eqref{eq:lag1-radius}. 

We next establish the uniform statement. For any
\(\theta\in\Theta_r^{(j)}\), let
\(
\Delta_\theta:=\theta^\star-\theta.
\)
From \eqref{eq:residual-map}, one has
\(
\widetilde W(\theta)
=
\widetilde W(\theta^\star)+\Delta_\theta\Phi.
\)
Substitution into the definition of \(\widehat\Sigma_1(\theta)\) gives
\(
\widehat\Sigma_1(\theta)
-\widehat\Sigma_1(\theta^\star)
=
\frac1T\big[
\Delta_\theta\Phi S_1
\widetilde W(\theta^\star)^\top
+
\widetilde W(\theta^\star)S_1
\Phi^\top\Delta_\theta^\top
+
\Delta_\theta\Phi S_1
\Phi^\top\Delta_\theta^\top
\big].
\)
Therefore,  \(\|S_1\|_2=1\) yields
\(
\big\|
\widehat\Sigma_1(\theta)
-\widehat\Sigma_1(\theta^\star)
\big\|_2
\le
\frac{2}{T}
\|\Delta_\theta\|_2
\|\Phi\|_2
\|\widetilde W(\theta^\star)\|_2\\
+
\frac1T
\|\Delta_\theta\|_2^2
\|\Phi\|_2^2 .
\)
By Lemma~\ref{lem:large-T-consistency},
\(
\sup_{\theta\in\Theta_r^{(j)}}
\|\Delta_\theta\|_2
=
O_{\mathbb P}\!\left(
\sqrt{{\log T}/{T}}
\right),
\)
while
\(
\|\Phi\|_2=O_{\mathbb P}(\sqrt T),
\qquad
\|\widetilde W(\theta^\star)\|_2
=O_{\mathbb P}(\sqrt T).
\)
Consequently, uniformly over \(\Theta_r^{(j)}\), the two terms above are
respectively
\(
O_{\mathbb P}\!\left(\sqrt{{\log T}/{T}}\right)
\,\,\text{and}\,\,
O_{\mathbb P}\!\left({\log T}/{T}\right),
\)
.
Finally, the triangle inequality gives
\(
\sup_{\theta\in\Theta_r^{(j)}}
\|\widehat\Sigma_1(\theta)-\Sigma_1\|_2
\le
\sup_{\theta\in\Theta_r^{(j)}}
\|\widehat\Sigma_1(\theta)-\widehat\Sigma_1(\theta^\star)\|_2
+
\|\widehat\Sigma_1(\theta^\star)-\Sigma_1\|_2,
\)
which proves \eqref{eq:lag1-uniform}.
\end{proof}

{Since $\Sigma_1=-VA^{\star\top}$, $V$ is identified only when $A^\star$ is nonsingular, which the certificate of $\Sigma_{\rm true}$ does not need. The refined set first gives consistent estimators of $V$ and $W$.}

\begin{corollary}
\label{cor:WV}
Under the conditions of Corollary~\ref{cor:cov-recovery}, let $A^\star$ be nonsingular, fix $j$, and for $\theta\in\Theta_r^{(j)}$ let $\widehat\Sigma_1^{(j)}(\theta):=\widehat\Sigma_1(\theta)$ and $\widehat A^{(j)}(\theta):=A(\theta)$. Then
\begin{align}
\widehat V^{(j)}(\theta)
&:=
\Pi_{\mathbb S_+^n}\big(
\sym(-\widehat\Sigma_1^{(j)}(\theta)
\widehat A^{(j)}(\theta)^{-\top})\big),
\nonumber\\
\widehat W^{(j)}(\theta) :=& \Pi_{\mathbb S_+^n}\big(\Sigma^{(j+1)} - \widehat V^{(j)}(\theta) - \widehat A^{(j)}(\theta)\widehat V^{(j)}(\theta)\widehat A^{(j)}(\theta)^\top\big)
\label{eq:WV-estimators}
\end{align}
satisfy, uniformly over \(\theta\in\Theta_r^{(j)}\),
\begin{align}
\|\widehat V^{(j)}(\theta)-V\|_2,\ \|\widehat W^{(j)}(\theta)-W\|_2
=O_{\mathbb P}(\sqrt{\log T/T}).
\label{eq:WV-rates}
\end{align}
\end{corollary}

\begin{proof}
\new{By Lemma~\ref{lem:lag1-radius}, \(\Sigma_1=-VA^{\star\top}\) and
\(\|\widehat\Sigma_1(\theta^\star)-\Sigma_1\|_2
=O_{\mathbb P}(\sqrt{\log T/T})\), while
\eqref{eq:lag1-uniform} extends this uniformly over the refined set, so
that
\(
\sup_{\theta\in\Theta_r^{(j)}}\|\widehat\Sigma_1^{(j)}(\theta)-\Sigma_1\|_2
=O_{\mathbb P}(\sqrt{\log T/T}).
\)}
By Lemma~\ref{lem:large-T-consistency},
\(\|\widehat A^{(j)}(\theta)-A^\star\|_2=O_{\mathbb P}(\sqrt{\log T/T})\)
uniformly, and since \(A^\star\) is nonsingular the inverse is
continuous in a neighbourhood, so
\(\|\widehat A^{(j)}(\theta)^{-\top}-(A^\star)^{-\top}\|_2\) obeys the same
rate. {Combining the two expansions gives, uniformly in $\theta$,
\[
\|-\widehat\Sigma_1^{(j)}(\theta)
\widehat A^{(j)}(\theta)^{-\top}-V\|_2
=O_{\mathbb P}(\sqrt{\log T/T}).
\]
Symmetrization is nonexpansive in spectral norm. Euclidean projection onto
$\mathbb S_+^n$ is nonexpansive in Frobenius norm. Using
$\|X\|_2\le\|X\|_F\le\sqrt n\|X\|_2$ shows that the spectral-norm rate is
preserved. This proves the first bound in \eqref{eq:WV-rates}.}
The second follows from
\(W=\Sigma_{\rm true}-V-A^\star V(A^\star)^\top\),
Corollary~\ref{cor:cov-recovery}, the first bound, and the same
Frobenius-projection argument.
\end{proof}

{For certification, the lag-one moment is bounded jointly with the zero-lag moment by the pivotal argument of Lemma~\ref{lem:cov-lower} in dimension $2n$, and the $A$-block of the cover keeps $A^\star$ away from singularity.}

\begin{lemma}
\label{lem:WV-certified}
Let Assumptions~\ref{noises}--\ref{ass:exogenous} hold, let
$\Sigma_{\rm true}\preceq\Sigma^{(j)}$, and let $A_{\rm ref}^{(j)}$ and $G_{A,s}^{(j)}$ be the $A$-blocks of $C_\theta^{(j)}$ and $G_{\theta,s}^{(j)}$ in \eqref{eq:cover-stacked}. With $M_\psi:=D_\psi\bar\Phi^{-1}$, $\widetilde W_c^{(j)}:=\widetilde W(C_\theta^{(j)})$, define
\begin{align}
\widehat\Sigma_{1,c}^{(j)}&:=T^{-1}\widetilde W_c^{(j)}S_1\widetilde W_c^{(j)\top},\
\Upsilon_1:=T^{-1}\Phi S_1\widetilde W_c^{(j)\top},\nonumber\\
\Upsilon_2&:=T^{-1}\widetilde W_c^{(j)}S_1\Phi^\top,\
\Upsilon_3:=T^{-1}\Phi S_1\Phi^\top,\nonumber\\
\varepsilon_1^{(j)}&:=\|C_j\|_2\big(\|M_\psi\Upsilon_1\|_2+\|\Upsilon_2M_\psi^\top\|_2\big)
+\|C_j\|_2^2\|M_\psi\Upsilon_3M_\psi^\top\|_2,\nonumber\\
r_A^{(j)}&:=\min\big\{\textstyle\sum_s\|G_{A,s}^{(j)}\|_2,\;
\|C_j\|_2\|M_\psi[I_n\;0]^\top\|_2\big\},\nonumber\\
\bar s_0^{(j)}&:=\beta_T\|\Sigma^{(j)}\|_2,\
\underline s_1^{(j)}:=[\sigma_{\min}(\widehat\Sigma_{1,c}^{(j)})-\varepsilon_1^{(j)}]_+,\nonumber\\
\bar s_1^{(j)}&:=\min\{\|\widehat\Sigma_{1,c}^{(j)}\|_2+
\varepsilon_1^{(j)},\bar s_0^{(j)}\}.
\label{eq:computable-lag-bounds}
\end{align}
With $T':=T-1$, the residue classes $\mathcal J_q$ of $\{0,\ldots,T-2\}$ modulo $3$, $T'_q:=|\mathcal J_q|\ge2n$, and $\delta_1\in(0,1)$, let
\begin{align*}
\alpha'&:=\textstyle\sum_{q}\tfrac{T'_q}{T'}
\ell_{2n,T'_q}(\tfrac{\delta_1}{6}),
&\beta'&:=\textstyle\sum_{q}\tfrac{T'_q}{T'}
 u_{2n,T'_q}(\tfrac{\delta_1}{6}),\\
{\bar c}&:=\tfrac12(\alpha'^{-1}+\beta'^{-1}),
&{\bar\rho}&:=\tfrac12(\alpha'^{-1}-\beta'^{-1}),\\
\bar\sigma_1^{(j)}&:=\tfrac{T}{T'}
({\bar c}\bar s_1^{(j)}+{\bar\rho}\bar s_0^{(j)}),
&\underline\sigma_1^{(j)}&:=\tfrac{T}{T'}
[{\bar c}\underline s_1^{(j)}-{\bar\rho}\bar s_0^{(j)}]_+,\\
\underline a_A^{(j)}&:=[\sigma_{\min}(A_{\rm ref}^{(j)})-r_A^{(j)}]_+,
&\bar a_A^{(j)}&:=\|A_{\rm ref}^{(j)}\|_2+r_A^{(j)}.
\end{align*}
If $\underline a_A^{(j)}>0$, then, with probability at least
$1-(\delta_c+\delta_s+\delta_1)$, $A^\star$ is nonsingular and
\begin{align}
& \underline v^{(j)}I_n\preceq V,
\quad V\preceq\Sigma^{(j)},
\quad V\preceq\bar v^{(j)}I_n,
\label{eq:V-upper-convex}\\
& W\preceq\bar W^{(j)}
:=\Sigma^{(j)}-\underline v^{(j)}
\big(1+(\underline a_A^{(j)})^2\big)I_n,
\label{eq:W-upper-convex}
\end{align}
where
$\bar v^{(j)}:=\bar\sigma_1^{(j)}/\underline a_A^{(j)}$ and
$\underline v^{(j)}:=\underline\sigma_1^{(j)}/\bar a_A^{(j)}$.
\end{lemma}

\begin{proof}
{Let $\theta\in\Theta_r^{(j)}$. Since
$\Theta_r^{(j)}\subseteq\Theta_{\rm sn}^{(j)}$ and $C_\theta^{(j)}=\theta_{\rm IV}$,
Lemma~\ref{lem:selfnormalized-MZ} gives
$\Delta_\theta:=C_\theta^{(j)}-\theta=C_j\Delta M_\psi$ with $\|\Delta\|_2\le1$.
{Here, $A_{\rm ref}^{(j)}$ is the $A$-block of the same cover center $C_\theta^{(j)}$.} Its first $n$ columns, together with
$\Theta_r^{(j)}\subseteq\Theta_{\rm pcz}^{(j)}$, give
$\|A(\theta)-A_{\rm ref}^{(j)}\|_2\le r_A^{(j)}$. Moreover,}
$\widetilde W(\theta)=\widetilde W_c^{(j)}+\Delta_\theta\Phi$, and hence
\begin{align*}
\widehat\Sigma_1(\theta)-\widehat\Sigma_{1,c}^{(j)}
={\Delta_\theta\Upsilon_1+\Upsilon_2\Delta_\theta^\top
+\Delta_\theta\Upsilon_3\Delta_\theta^\top ,}
\end{align*}
{whose norm is at most $\varepsilon_1^{(j)}$ after substituting
$\Delta_\theta=C_j\Delta M_\psi$.} Singular-value perturbation gives the first bound in
\eqref{eq:computable-lag-bounds}. Moreover, {on $\mathcal E_{\rm all}$, Lemma~\ref{lem:cov-lower} gives}
\[
\|\widehat\Sigma_1(\theta{^\star})\|_2
\le
\|{\widehat\Sigma(\theta^\star)}\|_2
\le {\beta_T\|\Sigma_{\rm true}\|_2\le}\bar s_0^{(j)},
\]
which proves the stated upper bound {at $\theta^\star$, the only point where it is used}.

Let $z_k:=\col(\widetilde w_k,\widetilde w_{k+1})$. Then
$z_k\sim\mathcal N(0,\Xi)$ with
\[
\Xi:=\begin{bmatrix}\Sigma_{\rm true}&\Sigma_1\\
\Sigma_1^\top&\Sigma_{\rm true}\end{bmatrix},
\qquad \Sigma_1=-VA^{\star\top},
\]
and samples within each $\mathcal J_q$ are independent. Writing
$z_k=\Xi^{1/2}g_k$ (no inverse of $\Xi$ is required) and repeating
Lemma~\ref{lem:cov-lower} with dimension $2n$ and three residue classes gives,
with probability at least $1-\delta_1$,
\[
\alpha'\Xi\preceq\widehat\Xi\preceq\beta'\Xi,
\qquad
\widehat\Xi:=\tfrac1{T'}\sum_{k=0}^{T-2}z_kz_k^\top.
\]
Thus, $D:=\Xi-{\bar c}\widehat\Xi$ satisfies
$-{\bar\rho}\widehat\Xi\preceq D\preceq{\bar\rho}\widehat\Xi$. For
$u=\col(x,y)$ and $v=\col(x,-y)$,
\[
4x^\top D_{12}y\le2{\bar\rho}
(x^\top\widehat\Xi_{11}x+y^\top\widehat\Xi_{22}y).
\]
Replacing $(x,y)$ by $(tx,y/t)$ and minimizing over $t>0$ yields
$\|D_{12}\|_2\le {\bar\rho}\|\widehat\Xi_{11}\|_2^{1/2}
\|\widehat\Xi_{22}\|_2^{1/2}$. Since
$\widehat\Xi_{11},\widehat\Xi_{22}\preceq
T'^{-1}\widetilde W^\star\widetilde W^{\star\top}$ and
$\widehat\Xi_{12}=(T/T')\widehat\Sigma_1(\theta^\star)$, while
$\theta^\star\in\Theta_r^{(j)}$ on $\mathcal E_{\rm all}$,
\[
\|\Sigma_1- {\bar c}\tfrac{T}{T'}\widehat\Sigma_1(\theta^\star)\|_2
\le {\bar\rho}\tfrac{T}{T'}\bar s_0^{(j)}.
\]
Weyl's inequality gives
$\underline\sigma_1^{(j)}\le\sigma_{\min}(\Sigma_1)$ and
$\|\Sigma_1\|_2\le\bar\sigma_1^{(j)}$. The generator bound likewise gives
$\underline a_A^{(j)}\le\sigma_{\min}(A^\star)$ and
$\|A^\star\|_2\le\bar a_A^{(j)}$.

Finally, $V=-\Sigma_1A^{\star-\top}$ gives
$\|V\|_2\le\bar v^{(j)}$, while
$\Sigma_1=-VA^{\star\top}$ gives
$\lambda_{\min}(V)\ge\underline v^{(j)}$. Also
$V\preceq\Sigma_{\rm true}\preceq\Sigma^{(j)}$. Hence
$V+A^\star VA^{\star\top}\succeq
\underline v^{(j)}(1+(\underline a_A^{(j)})^2)I_n$. Subtracting this from
$\Sigma_{\rm true}\preceq\Sigma^{(j)}$ proves
\eqref{eq:W-upper-convex}. The probability follows by a union bound.
\end{proof}

\subsection{{Consistency Test and Self-Consistent Set}}
\label{sec:consistency}
\begin{remark}
\label{rem:aggregate}
Lemma~\ref{lem:cov-lower} also gives $\widehat\Sigma(\theta^\star)\preceq\beta_T\Sigma_{\rm true}\preceq\beta_T\Sigma^{(j)}$, which could be added to \eqref{eq:statistical-noise-set} as a constraint on the full empirical covariance. We omit it because it cannot change any accepted certificate. Let $g(\theta):=\lammax((\Sigma^{(j)})^{-1/2}\widehat\Sigma(\theta)(\Sigma^{(j)})^{-1/2})$, which is continuous, and $\Theta'_r:=\Theta_r^{(j)}\cap\{g\le\beta_T\}$. Since $\delta_c/4<1/2$ and $\lammin\le\lammax$ with continuous laws, $\ell_{n,t}(\delta_c/4)<u_{n,t}(\delta_c/4)$, so $\alpha_T<\beta_T$. If an update is accepted over a nonempty $\Theta'_r$, then $g\le\alpha_T$ there, were $g(\theta_1)>\beta_T$ for some $\theta_1\in\Theta_r^{(j)}$, the segment from a point of $\Theta'_r$ to $\theta_1$, which lies in the convex $\Theta_r^{(j)}$, would contain a point with $g=\beta_T>\alpha_T$ in $\Theta'_r$. Hence, $\Theta'_r=\Theta_r^{(j)}$, and after an accepted update $\widehat\Sigma\preceq\alpha_T\Sigma^{(j+1)}\prec\beta_T\Sigma^{(j+1)}$ on all later sets. The bound only provides the sanity check $R_j:=\beta_T\Sigma^{(j)}-\widehat\Sigma_{\rm LS}\succeq0$, which holds on $\mathcal E_{\rm all}$ because $R_j\succeq\widehat\Sigma(\theta^\star)-\widehat\Sigma_{\rm LS}$ by \eqref{eq:residual-identity}.
\end{remark}

A single instrument always contains its own estimate, $\theta_{\rm IV}^{(i)}\in\Theta_{\rm sn}^{(i,j)}$, so it cannot reveal that the data contradict the model class. Two or more instruments can: under a correct model they all contain $\theta^\star$, whereas under misspecification their estimates converge to different limits and their sets separate.
\begin{theorem}
\label{thm:consistency-test}
Reject the model class \eqref{eq:syst}, \eqref{eq:CMZ-prior} if, at some
iteration $j$, $\Theta_r^{(j)}=\emptyset$, $\Theta_{\rm pcz}^{(j)}=\emptyset$, $\overline h_r<d_r-s_r$ for some cut, or $R_j\not\succeq0$.
\begin{enumerate}
\item[(i)] Under Assumptions~\ref{noises}--\ref{ass:exogenous}, a false
rejection has probability at most $\delta_c+\delta_s$, simultaneously over
all iterations and outer sets.
\item[(ii)] Let {$N_\psi\ge2$} and suppose, without assuming \eqref{eq:syst} or
Assumptions~\ref{noises}--\ref{ass:exogenous}, that
$\|\Sigma^{(0)}\|_2=O_{\mathbb P}(1)$, that the convergences of
Assumption~\ref{ass:asymptotic-informativity} hold for instruments $1$ and
$2$, and that $\theta_{\rm IV}^{(1)}-\theta_{\rm IV}^{(2)}\to c\neq0$ in
probability. Then the probability of rejection tends to one.
\end{enumerate}
\end{theorem}

\begin{proof}
(i) On $\mathcal E_{\rm all}$, $\theta^\star\in\Theta_r^{(j)}$ for all $j$
by Lemma~\ref{lem:finite-T-convergence} and
Theorem~\ref{thm:reduced-set}(iv), and $\Theta_r^{(j)}$ lies in every outer
set. Also,
$R_j\succeq0$ {by Remark~\ref{rem:aggregate}}, and
$\overline h_r\ge\langle L_r,\theta^\star\rangle\ge d_r-s_r$. A false rejection therefore implies
$\mathcal E_{\rm all}^c$. (ii) As in
\eqref{eq:sn-radius}, $\Theta_{\rm sn}^{(i,0)}$ lies in the ball of radius
$\|\Sigma^{(0)}\|_2^{1/2}\|D_\psi^{(i)}\|_2\|(\bar\Phi^{(i)})^{-1}\|_2
=O_{\mathbb P}(\sqrt{\log T/T})$ about $\theta_{\rm IV}^{(i)}$, by the
argument of Lemma~\ref{lem:large-T-consistency}, which uses only the stated
convergences. Since the centers stay near distance $\|c\|_2>0$, the balls,
and hence $\Theta_r^{(0)}$, are eventually disjoint.
\end{proof}

{The test detects misspecifications that yield distinct limiting IV estimates, but not those for which these limits coincide. Checking $\Theta_r^{(j)}=\emptyset$ is one feasibility problem with $N_\psi$ LMIs of size $2n+m$. Under correlated noise, $y_{k-1}$ is no longer a valid instrument while past inputs remain valid in open loop (Remark~\ref{rem:closed-loop-data}), so the IV estimates separate as in part~(ii). Finally, pairing each model with the covariance implied by its own realized noise, instead of one certificate for all models, removes the recursion and the initial certificate.}

\begin{proposition}
\label{prop:self-consistent}
For an instrument (index omitted), let
$S:=\bar\Phi D_\psi^{-2}\bar\Phi^\top$ and
\begin{align}
\Theta_{\rm sc}:=\big\{\theta:\ \alpha_T(\theta_{\rm IV}-\theta)S
(\theta_{\rm IV}-\theta)^\top\preceq\widehat\Sigma(\theta)\big\}.
\label{eq:self-consistent}
\end{align}
\begin{enumerate}
\item[(i)] On $\mathcal E_{\rm all}$, $\theta^\star\in\Theta_{\rm sc}$.
\item[(ii)] $K_S:=\alpha_TS-Q_T\succ0$ if and only if
$\lammax(M_q^\top M_q)<\alpha_T$, with $M_q$ as in
Remark~\ref{rem:sn-envelope}. Then, with
$\theta_c:=(\alpha_T\theta_{\rm IV}S-\theta_{\rm LS}Q_T)K_S^{-1}$ and
$R_{\rm sc}:=\widehat\Sigma_{\rm LS}+\theta_cK_S\theta_c^\top
-\alpha_T\theta_{\rm IV}S\theta_{\rm IV}^\top
+\theta_{\rm LS}Q_T\theta_{\rm LS}^\top$, $\Theta_{\rm sc}$ is empty if
$R_{\rm sc}\not\succeq0$ and equals
$\mathcal B(\theta_c;R_{\rm sc}^{1/2},K_S^{-1/2})$ otherwise.
\item[(iii)] In that case, with $M_c:=K_S^{-1/2}Q_T^{1/2}$,
$\Omega_c:=(\theta_{\rm LS}-\theta_c)Q_T^{1/2}$ and any
$\tau>\lammax(M_c^\top M_c)$,
\begin{align}
& \Sigma_{\rm sc}:= \nonumber \\ & \alpha_T^{-1}\big[\widehat\Sigma_{\rm LS}
+\tau R_{\rm sc}+\Omega_c(I_{n+m}-\tau^{-1}M_c^\top M_c)^{-1}\Omega_c^\top\big]
\label{eq:Sigma-sc}
\end{align}
satisfies $\Sigma_{\rm true}\preceq\Sigma_{\rm sc}$ on $\mathcal E_{\rm all}$.
\end{enumerate}
\end{proposition}

\begin{proof}
(i) The proof of Lemma~\ref{lem:iv} gives
$\bar N=\Sigma_{\rm true}^{1/2}\Delta^\star D_\psi$ with
$\|\Delta^\star\|_2\le1$, so
$\bar ND_\psi^{-2}\bar N^\top\preceq\Sigma_{\rm true}$. Combine with
$\bar N=(\theta_{\rm IV}-\theta^\star)\bar\Phi$ and
$\alpha_T\Sigma_{\rm true}\preceq\widehat\Sigma(\theta^\star)$
(Lemma~\ref{lem:cov-lower}). (ii) Since
$M_q^\top M_q=Q_T^{1/2}S^{-1}Q_T^{1/2}$, $K_S\succ0$ if and only if
$\lammax(Q_T^{1/2}S^{-1}Q_T^{1/2})<\alpha_T$. By
Lemma~\ref{lem:residual-identity}(ii), \eqref{eq:self-consistent} reads
$\alpha_T(\theta_{\rm IV}-\theta)S(\theta_{\rm IV}-\theta)^\top
-(\theta_{\rm LS}-\theta)Q_T(\theta_{\rm LS}-\theta)^\top\preceq
\widehat\Sigma_{\rm LS}$. Completing the square gives
$(\theta-\theta_c)K_S(\theta-\theta_c)^\top\preceq R_{\rm sc}$, and the
factorization in the proof of Lemma~\ref{lem:cover}(iv) gives the form.
(iii) On $\Theta_{\rm sc}$,
$(\theta_{\rm LS}-\theta)Q_T^{1/2}=\Omega_c+R_{\rm sc}^{1/2}\Delta M_c$ with
$\|\Delta\|_2\le1$, the full-block uncertainty of
Remark~\ref{rem:sn-envelope} with $(\Omega_q,C_j,M_q)$ replaced by
$(\Omega_c,R_{\rm sc}^{1/2},M_c)$. Hence
$\alpha_T\Sigma_{\rm true}\preceq\widehat\Sigma(\theta^\star)\preceq
\alpha_T\Sigma_{\rm sc}$.
\end{proof}

{$\Theta_{\rm sc}$ is a recursion-free confidence set that needs no initial certificate. It may be intersected with $\Theta_r^{(j)}$ without affecting containment, and neither set generally contains the other.}
Replacing $\widehat\Sigma(\theta)/\alpha_T$ in \eqref{eq:self-consistent} by
one certificate recovers $\Theta_{\rm sn}^{(j)}$, and \eqref{eq:one-shot} is
the corresponding fixed point. In the low-data regime $K_S$ is indefinite,
$\Theta_{\rm sc}$ is nonconvex, and Algorithm~\ref{alg:iterative-refinement}
with the prior remains the tool.

\section{Simulation Results}
\label{sec:simulations}
\begingroup
We ask (Q1) whether the guarantees hold, (Q2) how conservative the
certificates are as $T$ grows, (Q3) what a second instrument adds, and (Q4)
whether the test of Theorem~\ref{thm:consistency-test} detects a wrong model
class.

\begin{figure*}[t]
\centering
\includegraphics[width=\textwidth]{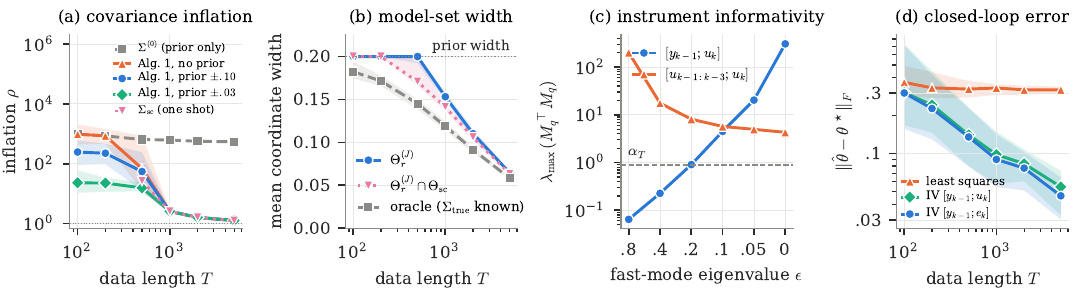}
\caption{Medians (lines) and $10$--$90\%$ ranges (bands). (a) Inflation
$\rho$ of the covariance certificates, single instrument, for priors of
half-width $.03$, $.10$ and $.30$. (b) Width $\bar w$
of the output set $\Theta_r^{(J)}$, of $\Theta_r^{(J)}\cap\Theta_{\rm sc}$,
and of the oracle set; the output equals the prior box until the record is
informative. (c) Informativity $\lambda_q$ of each instrument alone at
$T=5000$ when the last row of $A^\star$ is $[0\;0\;\epsilon]$; values below
$\alpha_T$ (dashed) are informative. (d) Closed-loop error
$\|\hat\theta-\theta^\star\|_F$; $[y_{k-1};u_k]$ is an invalid instrument
under feedback.}
\label{fig:sim-summary}
\end{figure*}

We consider the system (1) with 3 states and
\begin{align*}
A^\star&=\begin{bmatrix}.90&.20&0\\0&.85&.20\\.04&0&.80\end{bmatrix},
& B^\star&=\begin{bmatrix}0&.5&1\end{bmatrix}^{\!\top},
\end{align*}
We also assume $W=.02I_3$, $V=.002I_3$, and $u_k\sim\mathcal N(0,1)$, unless stated
otherwise. In this and the following subsections, the prior is the box
$\mathcal M_{\theta,p}=\{\theta:|\theta_{ab}-(C_{\theta,p})_{ab}|\le h\}$ of
half-width $h$, i.e., $G_{\theta,p,i}=hE_i$ in \eqref{eq:CMZ-prior}. Its
center $C_{\theta,p}$ is $\theta^\star$ plus independent offsets drawn
uniformly on $[-.4h,.4h]$ for each trajectory, so the box contains
$\theta^\star$ but is not centered on it. We use $h=.10$ unless stated
otherwise. The instruments are
$\psi_k^{(1)}=[y_{k-1};u_k]$ and the past-input instrument
$\psi_k^{(2)}=[u_{k-1};u_{k-2};u_{k-3};u_k]$ of
Remark~\ref{rem:closed-loop-data}. We use $\delta_c=\delta_s=\delta_1=.01$,
so the joint guarantees hold with probability at least $.98$, with
$\delta_{s,1}=.9\delta_s$ and $\delta_{s,2}=.1\delta_s$ when $N_\psi=2$. The regularization in $V_\psi^{(i)}$ of \eqref{eq:Dpsi} (Lemma~\ref{lem:iv})
is $\lambda_\psi=.1T$, and the trace weight in \eqref{eq:reduced-cov-SDP},
also used to rank the candidates in Algorithm~\ref{alg:iterative-refinement},
is $H=I_3$. The quantiles $\ell_{n,t}$ and $u_{n,t}$ in
\eqref{eq:pivotal-quantiles}, which give $\alpha_T$ and $\beta_T$ in
\eqref{eq:alpha-beta-exact} (Lemma~\ref{lem:cov-lower}), are estimated from
$2\times10^6$ Monte Carlo draws with a four-standard-deviation
order-statistic margin (Remark~\ref{rem:pivotal-quantiles}).

Algorithm~\ref{alg:iterative-refinement} starts
from $\Sigma^{(0)}$ (Lemma~\ref{lem:init-cert-prior}) and accepts at each
iteration the smallest-trace admissible candidate among
\eqref{eq:reduced-cov-SDP}, \eqref{eq:Gamma-tau} and, if $N_\psi=2$,
\eqref{eq:sproc}. It stops after $J$ iterations, when the trace decreases by
less than $1\%$. Its outputs, the model set $\Theta_r^{(J)}$ and the
certificate $\Sigma^{(J)}$, are the inputs to Part~II~\cite{PartII}. The {single-instrument} configuration ($N_\psi=1$) uses only
$\psi^{(1)}$. Then, $\Theta_r^{(j)}$ in \eqref{eq:reduced-set} has one IV
constraint, and its cover is the interval hull of
$\Theta_{\rm sn}^{(j)}$ (Lemma~\ref{lem:cover}(i)), intersected exactly with
the prior (Lemma~\ref{lem:full-cover}), without cuts. The candidates are
\eqref{eq:reduced-cov-SDP} and \eqref{eq:Gamma-tau}. The
{two-instrument} configuration ($N_\psi=2$) adds $\psi^{(2)}$, with
$\delta_s$ split as $\delta_{s,1}=.9\delta_s$, $\delta_{s,2}=.1\delta_s$.
Then, $\Theta_r^{(j)}$ has both IV constraints, and the cover is tightened
by the closed-form cuts of Lemma~\ref{lem:cover}(iv) along $\pm E_s$, which
intersect it with the interval hull of $\Theta_{\rm sn}^{(2,j)}$ at no SDP
cost. The candidates are \eqref{eq:reduced-cov-SDP}, \eqref{eq:Gamma-tau}
for each instrument, and the S-procedure candidate \eqref{eq:sproc} over
$\Theta_{\rm sn}^{(1,j)}\cap\Theta_{\rm sn}^{(2,j)}$.

We use the following measures for evaluation. The {inflation}
$\rho(\Sigma):=\lammax(\Sigma_{\rm true}^{-1/2}\Sigma\Sigma_{\rm true}^{-1/2})\ge1$
is the worst-direction factor by which a valid certificate overestimates
$\Sigma_{\rm true}$. The {width}
$\bar w(\Theta):=\frac1{n(n+m)}\sum_{a,b}(\max_{\Theta}\theta_{ab}-\min_{\Theta}\theta_{ab})$
is the mean side of the bounding box of $\Theta$ ($.20$ for the prior). The {oracle width} $\bar w^\star$ is the width of
\eqref{eq:reduced-set} built with $C_j=\Sigma_{\rm true}^{1/2}$, so
$\bar w-\bar w^\star$ is the cost of not knowing the covariance. A record is
{informative} if $\lambda_q:=\lammax(M_q^\top M_q)<\alpha_T$
(Remark~\ref{rem:sn-envelope}), which is exactly when \eqref{eq:Gamma-tau},
\eqref{eq:one-shot} and $\Theta_{\rm sc}$ exist. Values are medians.

\subsection{Verification of the Guarantees (Q1)}
\label{sec:sim-verify}
We show that the true model and the true covariance are
never removed: on an event of probability at least
$1-\delta_c-\delta_s=.98$, $\theta^\star\in\Theta_r^{(J)}$ and
$\Sigma_{\rm true}\preceq\Sigma^{(J)}$ (Theorem~\ref{thm:reduced-set} and
Lemma~\ref{lem:finite-T-convergence}). We generated $100$ independent
trajectories for each $T\in\{100,200,500,1000,2000,5000\}$, ran
Algorithm~\ref{alg:iterative-refinement} on each of them in both
configurations, and counted the runs in which these two properties hold.
Table~\ref{tab:verify} shows that both held in every run, i.e., with
empirical frequency $1.00$ against the guaranteed $.98$. The margin reflects
the conservatism of the confidence bounds of Lemmas~\ref{lem:cov-lower}
and~\ref{lem:iv}. 

\begin{table}[t]
\centering
\caption{Containment of the true model and covariance.}
\label{tab:verify}
\scriptsize
\setlength{\tabcolsep}{4pt}
\begin{tabular}{@{}rcccc@{}}
\toprule
& \multicolumn{2}{c}{$\theta^\star\in\Theta_r^{(J)}$} & \multicolumn{2}{c}{$\Sigma_{\rm true}\preceq\Sigma^{(J)}$}\\
\cmidrule(lr){2-3}\cmidrule(l){4-5}
$T$ & $N_\psi{=}1$ & $N_\psi{=}2$ & $N_\psi{=}1$ & $N_\psi{=}2$\\
\midrule
$100$ & $1.00$ & $1.00$ & $1.00$ & $1.00$\\
$200$ & $1.00$ & $1.00$ & $1.00$ & $1.00$\\
$500$ & $1.00$ & $1.00$ & $1.00$ & $1.00$\\
$1000$ & $1.00$ & $1.00$ & $1.00$ & $1.00$\\
$2000$ & $1.00$ & $1.00$ & $1.00$ & $1.00$\\
$5000$ & $1.00$ & $1.00$ & $1.00$ & $1.00$\\
\bottomrule
\end{tabular}
\end{table}

\subsection{Tightness Versus Data Length (Q2)}
\label{sec:sim-tight}
Without an informative record, only the prior and
\eqref{eq:reduced-cov-SDP} can refine $\Sigma^{(0)}$
(Remark~\ref{rem:sn-envelope}). With one,
Lemma~\ref{lem:large-T-consistency} and Corollary~\ref{cor:cov-recovery}
predict convergence at the rate $\sqrt{\log T/T}$. Table~\ref{tab:cov} and
Fig.~\ref{fig:sim-summary}(a,b) show both regimes for the single-instrument
configuration with $h=.10$, on the trajectories of Table~\ref{tab:verify}.

\begin{table}[t]
\centering
\caption{Single instrument, prior half-width $h=.10$, medians over $100$
trajectories per $T$. $p_{\rm inf}$: fraction with $\lambda_q<\alpha_T$;
$J$: iterations. Inflation $\rho$ of $\Sigma^{(0)}$; of
Algorithm~\ref{alg:iterative-refinement} without the prior (candidate
\eqref{eq:Gamma-tau} only); of its output $\Sigma^{(J)}$; of
\eqref{eq:one-shot}; and of $\Sigma_{\rm sc}$
(Proposition~\ref{prop:self-consistent}). Width of $\Theta_r^{(J)}$, of
$\Theta_r^{(J)}\cap\Theta_{\rm sc}$, and of the oracle set. --: no
trajectory is informative.}
\label{tab:cov}
\scriptsize
\setlength{\tabcolsep}{1.9pt}
\begin{tabular}{@{}rccr|rrrrr|ccc@{}}
\toprule
& & & & \multicolumn{5}{c|}{inflation $\rho$} & \multicolumn{3}{c}{width $\bar w$}\\
$T$ & $\alpha_T$ & $p_{\rm inf}$ & $J$ & $\Sigma^{(0)}$ & no prior & $\Sigma^{(J)}$ &
\eqref{eq:one-shot} & $\Sigma_{\rm sc}$ & $\Theta_r^{(J)}$ & $\cap\Theta_{\rm sc}$ & oracle\\
\midrule
$100$ & $.40$ & $0$ & $2$ & $973$ & $973$ & $246$ & -- & -- & $.200$ & $.200$ & $.182$\\
$200$ & $.55$ & $0$ & $2$ & $853$ & $853$ & $224$ & -- & -- & $.200$ & $.200$ & $.172$\\
$500$ & $.69$ & $.71$ & $32$ & $642$ & $69.2$ & $54.7$ & $26.6^\dagger$ & $26.4^\dagger$ & $.200$ & $.171$ & $.145$\\
$1000$ & $.78$ & $1$ & $12$ & $620$ & $2.56$ & $2.56$ & $2.54$ & $2.53$ & $.153$ & $.141$ & $.119$\\
$2000$ & $.84$ & $1$ & $8$ & $562$ & $1.60$ & $1.60$ & $1.60$ & $1.60$ & $.110$ & $.106$ & $.091$\\
$5000$ & $.90$ & $1$ & $6$ & $542$ & $1.26$ & $1.26$ & $1.25$ & $1.25$ & $.065$ & $.063$ & $.058$\\
\bottomrule
\multicolumn{12}{@{}l}{$^\dagger$Median over the $71$ informative trajectories.}
\end{tabular}
\end{table}

For $T\le200$, no record is informative. This is expected: the radius of
$\Theta_{\rm sn}^{(j)}$ in \eqref{eq:sn-radius} scales with
$\|D_\psi\|_2=O(\sqrt{\log T/T})$ and is wider than the prior box, while
$\alpha_T$ is small ($.40$ at $T=100$), so
$\lammax(M_q^\top M_q)\ge\alpha_T$ and \eqref{eq:Gamma-tau} is not
admissible (Remark~\ref{rem:sn-envelope}). The recursion without the prior
therefore cannot improve on $\Sigma^{(0)}$. The prior and the SDP lower
$\rho$ by a factor of $4$, but $\Theta_r^{(J)}$ stays the prior box, so the
prior limits the result.
 At $T=500$,
$71\%$ of the records are informative. On these, \eqref{eq:one-shot}
reaches $\rho=26.6$, whereas the iteration stops at $54.7$, because its
contraction factor $\tau/\alpha_T$ is near one. For $T\ge1000$,
Algorithm~\ref{alg:iterative-refinement}, \eqref{eq:one-shot} and
$\Sigma_{\rm sc}$ in \eqref{eq:Sigma-sc} agree within $1.2\%$. From $T=1000$ to $5000$, $\rho$
falls from $2.56$ to $1.26$, $\bar w/\bar w^\star$ falls from $1.29$ to
$1.11$, and $\bar w/\sqrt{\log T/T}$ stays within $[1.5,1.9]$, as
Lemma~\ref{lem:large-T-consistency} predicts. Intersecting with
$\Theta_{\rm sc}$ closes $53$--$19\%$ of the gap $\bar w-\bar w^\star$ for
$T=500$--$5000$. Every accepted update satisfied
$\Gamma_j^\star\preceq\beta_T\Sigma^{(j+1)}$, so the bound omitted in
Remark~\ref{rem:aggregate} would change no entry.
The results conclude that, with little data the prior and the SDP provide the only refinement, whereas with a long record the certificates approach their
known-covariance counterparts at the rate predicted by Lemma~\ref{lem:large-T-consistency}, independently of the prior.

To check the effect of prior knowledge, Table~\ref{tab:prior} repeats the same configuration
with a tighter prior ($h=.03$) and a looser one ($h=.30$), $20$
trajectories each. With little data, the prior sets the result: at
$T=100$, $\rho$ is $22$, $246$ and $1940$ for $h=.03$, $.10$ and $.30$, in
proportion to the prior-only certificate $\Sigma^{(0)}$, and
$\Theta_r^{(J)}$ is the prior box. From $T=1000$ on, $\rho$ is the same for
all three priors, because the closed-form candidates do not use the prior.
The model set still benefits from a prior that is narrower than the
self-normalized set: at $T=1000$, $\bar w$ is $.058$, $.153$ and $.201$.
Containment held in all $240$ runs.

\begin{table}[t]
\centering
\caption{Effect of the prior half-width $h$ (single instrument; medians
over $100$ trajectories for $h=.10$, as in Table~\ref{tab:cov}, and $20$
for $h=.03$ and $.30$). $\rho$: inflation of $\Sigma^{(0)}$ at $T=100$ and
of the output $\Sigma^{(J)}$; $\bar w$: width of $\Theta_r^{(J)}$. The prior
alone gives $\bar w=2h$.}
\label{tab:prior}
\scriptsize
\setlength{\tabcolsep}{2.6pt}
\begin{tabular}{@{}l|c|cccc|cccc@{}}
\toprule
& $\rho(\Sigma^{(0)})$ & \multicolumn{4}{c|}{$\rho(\Sigma^{(J)})$} & \multicolumn{4}{c}{$\bar w$}\\
$h$ & $T{=}100$ & $100$ & $500$ & $1000$ & $5000$ & $100$ & $500$ & $1000$ & $5000$\\
\midrule
$.03$ & $100$ & $22.4$ & $14.4$ & $2.58$ & $1.27$ & $.060$ & $.060$ & $.058$ & $.047$\\
$.10$ & $973$ & $246$ & $54.7$ & $2.56$ & $1.26$ & $.200$ & $.200$ & $.153$ & $.065$\\
$.30$ & $7620$ & $1940$ & $44.9$ & $2.58$ & $1.27$ & $.600$ & $.539$ & $.201$ & $.065$\\
\bottomrule
\end{tabular}
\end{table}

\subsection{Second Instrument and Consistency Test (Q3, Q4)}
\label{sec:sim-inst}
An added instrument preserves containment and can only shrink the set
(Theorem~\ref{thm:reduced-set}), most when it is informative in other
directions. In Fig.~\ref{fig:sim-summary}(c), the last row of $A^\star$ is
$[0\;0\;\epsilon]$. As $\epsilon\to0$, $\sigma_{\min}(\bar\Phi^{(1)})$ falls
from $.23$ to $.006$, and $\psi^{(1)}$ is not informative for
$\epsilon\le.2$. The past input is insensitive to $\epsilon$ but weaker on
the slow modes, and alone it is never informative at this $T$.
Table~\ref{tab:multi}, again with $h=.10$, shows that the second instrument
improves every entry. For the nominal system, $\rho$ falls from $133$ to
$4.86$ at $T=500$ and from $2.61$ to $1.97$ at $T=1000$. The gain comes from
the S-procedure candidate \eqref{eq:sproc}, which bounds the residual
covariance over the intersection of the two sets and was accepted at every
iteration after the first for $T\ge500$. For $\epsilon=.1$, the sets are
complementary: at $T=5000$, $\rho$ falls from $13.5$ to $1.20$ and $\bar w$
from $.155$ to $.047$, close to the oracle value $.044$.

\begin{table}[t]
\centering
\caption{Single versus two instruments, prior half-width $h=.10$, medians
over $20$ trajectories per row. $\epsilon=.1$: last row of $A^\star$ set to
$[0\;0\;.1]$. $\bar w$, $\bar w^\star$: widths of $\Theta_r^{(J)}$ and of
the oracle set; $\rho$: inflation of $\Sigma^{(J)}$. Containment held in all
$240$ runs.}
\label{tab:multi}
\scriptsize
\setlength{\tabcolsep}{2.4pt}
\begin{tabular}{@{}lrcccccc@{}}
\toprule
& & \multicolumn{2}{c}{$\bar w$} & \multicolumn{2}{c}{$\bar w^\star$} & \multicolumn{2}{c}{$\rho$}\\
\cmidrule(lr){3-4}\cmidrule(lr){5-6}\cmidrule(l){7-8}
system & $T$ & $N_\psi{=}1$ & $N_\psi{=}2$ & $N_\psi{=}1$ & $N_\psi{=}2$ & $N_\psi{=}1$ & $N_\psi{=}2$\\
\midrule
nominal & $500$ & $.200$ & $.181$ & $.148$ & $.143$ & $133$ & $4.86$\\
nominal & $1000$ & $.154$ & $.141$ & $.124$ & $.116$ & $2.61$ & $1.97$\\
nominal & $5000$ & $.064$ & $.058$ & $.058$ & $.054$ & $1.26$ & $1.21$\\
$\epsilon=.1$ & $500$ & $.200$ & $.190$ & $.146$ & $.135$ & $28.1$ & $3.51$\\
$\epsilon=.1$ & $1000$ & $.200$ & $.129$ & $.120$ & $.099$ & $22.5$ & $1.83$\\
$\epsilon=.1$ & $5000$ & $.155$ & $.047$ & $.083$ & $.044$ & $13.5$ & $1.20$\\
\bottomrule
\end{tabular}
\end{table}

For the test, the process noise is made AR(1),
$\omega_k=a\,\omega_{k-1}+\sqrt{1-a^2}\,\varepsilon_k$, so the white-noise
class is wrong for $a>0$: $\psi^{(1)}$ becomes invalid while $\psi^{(2)}$
remains valid. Over $20$ trajectories per case ($h=.10$), the class was
never rejected for $a=0$, in line with
Theorem~\ref{thm:consistency-test}(i). With $N_\psi=2$, every trajectory
was rejected for $a=.95$ at $T\ge5000$ and for $a=.8$ at $T=2\cdot10^4$, and
none at $T=1000$. With $N_\psi=1$, at most $5\%$ were rejected, so the power
predicted by Theorem~\ref{thm:consistency-test}(ii) comes from
overidentification.
Thus, a second instrument tightens both certificates most when it is informative in directions where the first is weak.
\vspace{-9pt}

\subsection{Closed Loop, Separate Noise Bounds, and Cost}
\label{sec:sim-other}
With $W=.002I_3$, $V=.02I_3$, $u_k=Ky_k+e_k$, $K=-.1[1\;1\;1]$,
$e_k\sim\mathcal N(0,.35^2)$, the instruments $[y_{k-1};e_k]$ and
$[e_{k-1};e_{k-2};e_{k-3};e_k]$ (Remark~\ref{rem:closed-loop-data}), and
$h=.10$, containment held on all $30$ trajectories for each
$T\in\{100,1000,5000\}$. The least-squares certificate
$\widehat\Sigma_{\rm LS}/\alpha_T$ failed on all $30$ at $T=5000$. At that
$T$, the least-squares error levels off at $.32$, while the IV error reaches
$.047$ (Fig.~\ref{fig:sim-summary}(d)). Separating $W$ from $V$
(Lemma~\ref{lem:WV-certified}) required $T=2\cdot10^4$. One solve of
\eqref{eq:reduced-cov-SDP} (Clarabel) takes $0.08$\,s for $n=3$, $m=1$, and
$11.5$\,s for $n=6$, $m=2$, independent of $T$.

\endgroup

\section{Conclusion}
{This paper developed a finite-sample uncertainty description from one noisy input--state trajectory when the dynamics and the process- and measurement-noise covariances are unknown. Delayed IVs remove the errors-in-variables bias, and trajectory, prior, and statistical consistency preserve the dependence between each admissible model and its reconstructed noise. The model, realized-noise, normalized-noise, and covariance certificates contain their true counterparts, are recursively nested, and converge at the delayed-IV rate. Overidentifying instruments tighten the model set and turn the certificates into a test of the model class, and a self-consistent set tightens the description in the data-rich regime without recursion. The simulations confirmed every guarantee and showed that the prior matters only with little data, while a long record is needed before the certificates approach their known-covariance counterparts. Part~II~\cite{PartII} uses these certificates for robust domain-randomized LQR synthesis.}

\bibliographystyle{IEEEtran}
\bibliography{refs1}

\begin{IEEEbiography}[{\includegraphics[width=1in,height=1in,clip,keepaspectratio]{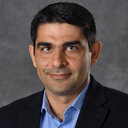}}]%
{\mdseries\normalfont Hamidreza Modares}
received the B.S., M.S., and Ph.D. degrees in electrical engineering from the University of Tehran, Shahrood University of Technology, and the University of Texas at Arlington, respectively. He is an Associate Professor at Michigan State University. His interests include safe learning-based control, multi-agent systems, and robotics.
\end{IEEEbiography}\vspace{-14pt}


\end{document}